\documentclass[aps,prb,twocolumn,superscriptaddress]{revtex4}

\usepackage[dvips]{graphicx}
\usepackage[dvips]{color}
\usepackage{amsmath}
\usepackage{amssymb}
\usepackage{multirow}

\def\e{\epsilon}

\def\t#1{\textrm{#1}}
\def\n{\nonumber \\}
\def\tensor{\otimes}

\def\tr#1#2{\tau_{#1}\rho_{#2}}

\begin{document}

\title{Topological classification with additional symmetries from Clifford algebras}

\author{Takahiro Morimoto}
\affiliation{Condensed Matter Theory Laboratory, RIKEN, Saitama, 351-0198, Japan}
\author{Akira Furusaki}
\affiliation{Condensed Matter Theory Laboratory, RIKEN, Saitama, 351-0198, Japan}
\affiliation{RIKEN Center for Emergent Matter Science (CEMS), Saitama, 351-0198, Japan}
\date{\today}

\begin{abstract}
We classify topological insulators and superconductors
in the presence of additional symmetries
such as reflection or mirror symmetries.
For each member of the 10 Altland-Zirnbauer symmetry classes,
we have a Clifford algebra defined by operators of the generic
(time-reversal, particle-hole, or chiral) symmetries
and additional symmetries, together with
gamma matrices in Dirac Hamiltonians
representing topological insulators
and superconductors.
Following Kitaev's approach, we classify gapped phases 
of non-interacting fermions under
additional symmetries by examining all possible distinct Dirac mass terms
which can be added to the set of generators of the Clifford algebra.
We find that imposing additional symmetries in effect changes
symmetry classes and causes shifts in the periodic table of topological
insulators and superconductors.
Our results are in agreement with the classification under reflection
symmetry recently reported by Chiu \textit{et al}.
Several examples are discussed including a topological crystalline insulator
with mirror Chern numbers and mirror superconductors.

\end{abstract}

\pacs{72.10.-d,73.20.-r,73.43.Cd}

\maketitle


\section{Introduction}
Since the theoretical proposals and experimental confirmations
of two- and three-dimensional topological insulators with
$\mathbb{Z}_2$ topological numbers,\cite{hasan-kane10,qi-zhang-rmp11,kane-mele05,kane2005z_,fu-kane-pump06,fu-kane-inv07,Bernevig15122006,konig2007quantum,hsieh2008topological,fu-kane-mele07,moore-balents07,roy-3dTI09}
topological characterization of gapped phases
has been intensively studied as a new way to classify states of matter,
beyond the conventional way in terms of broken symmetries.
The concept of topological insulators can be extended to any
system of non-interacting fermions with excitation gap
including superconductors
where fermionic quasiparticles are well described by the BCS mean-field theory.
More generally, the topological insulators and superconductors
can be defined as
systems of non-interacting (or weakly interacting) fermions
which have gapped excitation spectra in the bulk and
topologically stable gapless boundary excitations.
There exist a large variety of topological insulators and superconductors
in the broad sense defined above.
A prominent example of topological insulators is integer quantum Hall
states.\cite{TKNN}
Examples of topological superconductors include a one-dimensional
$p$-wave superconductor\cite{kitaev-uspekhi} and a chiral $p$-wave
superconductor in two dimensions.\cite{read-green}

The zoo of topological insulators and superconductors
has been classified theoretically.\cite{schnyder-ryu08,schnyder-aip-proc09,ryu-njp10,kitaev09,stone2011symmetries,abramovici-kalugin,wen-SPT-fermion12}
Systems of non-interacting fermions are known to be divided into
ten Altland-Zirnbauer (AZ) symmetry classes,\cite{AZ-classes}
in terms of the presence or absence of time reversal symmetry (TRS),
particle-hole symmetry (PHS), and chiral symmetry (or sublattice symmetry).
According to the
classification,\cite{schnyder-ryu08,schnyder-aip-proc09,ryu-njp10,kitaev09}
in every spatial dimension, there exist five distinct classes of topological
insulators and superconductors out of the ten AZ symmetry classes.
Among the five classes of topological insulators and superconductors,
three are characterized by an integer ($\mathbb{Z}$) topological index,
and two characterized by a binary ($\mathbb{Z}_2$) topological index.
Topological indices are defined from Bloch wave functions or Hamiltonian.
The general classification of topological insulators and superconductors
has been achieved in various ways:
stability analysis of gapless boundary states
against (random) perturbations,\cite{schnyder-ryu08,schnyder-aip-proc09}
dimensional reduction in representative massive
Dirac Hamiltonians,\cite{ryu-njp10}
and application of K-theory and Clifford
algebras.\cite{kitaev09,stone2011symmetries}
The last approach is most elegant and mathematically powerful.


The so-called periodic table of topological insulators and superconductors
was obtained by Kitaev\cite{kitaev09}
using K-theory and Clifford algebras.
In this approach, a Clifford algebra is formed from
generic symmetry transformations
such as TRS, PHS, or chiral symmetry.
One can then ask how many different types of generators
(related to Hamiltonian after spectral flattening) can be
added to the set of generators of the Clifford algebra.
The answer to this question is provided by ``classifying space''.
This formulation naturally leads to the periodic table of
topological insulators and superconductors for the 10 AZ symmetry classes
in any spatial dimension, which has the periodic structure
of period 2 or 8 coming from the Bott periodicity.

As an attempt to find a novel class of topological insulators in non-interacting systems,
Fu introduced the idea of topological crystalline insulators,\cite{fu-tci11}
i.e., band insulators which become topologically nontrivial only
when some crystalline symmetry is present in addition to
TRS.\cite{fu-tci11,zaanen2012space}
Furthermore, Fu and his collaborators made a theoretical
proposal\cite{hsieh12}
that SnTe should be a topological crystalline insulator whose
topological stability is guaranteed
by a mirror Chern number\cite{teo-fu-kane08}
defined on a mirror-invariant plane in the Brillouin zone.
Subsequent experimental studies
have confirmed that
SnTe and its alloys Sn$_{1-x}$Pb$_x$Te are topological crystalline
insulators.\cite{xu2012observation,tanaka2012experimental,dziawa2012topological}
The successful discovery of topological crystalline insulators
urges us to generalize the classification theory of topological
insulators and superconductors to include crystalline symmetries.
Indeed, Chiu \textit{et al.}\ have recently developed such classification
by explicitly constructing possible topological invariants for
representative Dirac Hamiltonians with a mirror symmetry
for each symmetry class.\cite{chiu13}


In this paper we follow the approach pioneered by Kitaev\cite{kitaev09}
to classify topological insulators and superconductors with
additional symmetries such as reflection symmetries.
Our approach is complementary to the one taken by Chiu \textit{et al.}\ 
and takes advantage of simple and powerful mathematics
of representation theory of Clifford algebras and K-theory.\cite{karoubi}
A drawback of our approach is that it does not give us
an explicit formula of topological invariants.

While we focus on non-interacting systems in this paper,
we note that new topological phases can appear in interacting systems,
since interactions can modify the above-mentioned classifications of
non-interacting fermions.
For one-dimensional systems of interacting particles,
modifications of the classification are explicitly shown and
the full classification is obtained, e.g., in terms of matrix product states.\cite{fidkowski-kitaev10,fidkowski-kitaev11,turner-pollman-berg11,chen-gu-wen11}
For bosonic systems, classification of symmetry protected topological
(SPT) phases in higher dimensions 
is recently proposed using group cohomology\cite{chen-gu-liu-wen12}
and (2+1)-dimensional Chern-Simons theory.\cite{lu-vishwanath12}

This paper is organized as follows.
In Sec.~\ref{sec:formalism} and \ref{sec:AZ} we briefly review Clifford
algebras and its application to the classification of topological insulators
and superconductors in zero dimension for the 10 AZ symmetry classes.
These sections give a summary of the theoretical formalism which is used
in the following sections.
In Sec.~\ref{sec:additional} we apply the formalism to classify
zero-dimensional topological insulators and superconductors
in the presence of an additional symmetry constraint.
We find that the additional symmetry in effect
shifts symmetry classes in the periodic table.
In Sec.~\ref{sec:dimensional shift} we take gamma matrices in Dirac
Hamiltonian as additional symmetry generators, to obtain
classification of $d$-dimensional topological insulators and superconductors
from classification at $d=0$.
In Sec.~\ref{sec:reflection} we derive the periodic table of topological
insulators/superconductors with a reflection symmetry.
Our table is in agreement with the one obtained earlier by Chiu \textit{et al.}
In Sec.~\ref{sec:multiple} we study cases when multiple additional
symmetries are imposed.
We find more complicated shuffling of symmetry classes in the periodic table.
In Sec.~\ref{sec:examples} we discuss several examples of topological
insulators and superconductors which are protected by reflection symmetries,
including topological crystalline insulators with mirror Chern numbers
or $\mathbb{Z}_2$ indices.
Some mathematical details are summarized in Appendices.

\section{Formalism\label{sec:formalism}}
We briefly introduce AZ symmetry classes and
Clifford algebras and summarize
our program for classifying topological insulators and superconductors
with additional symmetries in terms of real and complex Clifford algebras.

\subsection{Ten symmetry classes}
The AZ symmetry classes give classification of Hamiltonians
of free fermion systems.
Hamiltonians can be block diagonalized when they commute with
a unitary matrix representing (continuous) symmetry
transformation such as (spin) rotation or translation.
In the following discussions we assume that
Hamiltonians are already block diagonalized.
The Hamiltonians may still have discrete symmetries represented by
anti-unitary operators, and are
classified into the 10 AZ symmetry classes
according to the presence or absence of time-reversal and
particle-hole symmetries (Table~\ref{tab-bott}).\cite{AZ-classes}

When Hamiltonian $H$ has neither time-reversal nor
particle-hole symmetry, $H$ is in class A or AIII
of the complex classes; see Table~\ref{tab-bott}(a).\cite{AZ-classes,ryu-njp10}
When there exists a unitary transformation $\Gamma$ that changes the sign
of $H$ (i.e., $\Gamma^{-1}H\Gamma=-H$),
the Hamiltonian $H$ has so-called chiral symmetry and is a member
of class AIII.

Both time-reversal operator $T$ and particle-hole operator $C$
are an anti-unitary operator whose square
equals either plus or minus identity operator.
When $H$ has either time-reversal or particle-hole symmetry (or both),
$H$ belongs to real classes which are further divided into
eight symmetry classes listed in Table~\ref{tab-bott}(b),
in terms of the sign of $T^2$ and $C^2$.\cite{AZ-classes,ryu-njp10}
When $H$ has both $T$ and $C$ symmetries, the product of the two
symmetry operations yields a chiral symmetry transformation,
a unitary operator that anticommutes with $H$.

\begin{table}[tb]
\begin{center}
\caption{
Altland-Zirnbauer symmetry classes for (a) complex and (b) real cases.
The presence of time-reversal symmetry (TRS) and 
particle-hole symmetry (PHS) is denoted $+1$ or $-1$,
depending on whether they square to $+1$ or $-1$.
The absence of the TRS or PHS symmetry is denoted by ``$0$".
The next to last columns show
the classifying spaces (a) $C_q$ and (b) $R_q$ that
characterize zero-dimensional Hamiltonian
in each symmetry class.
The last columns denote type of possible topological numbers,
i.e., the number of disconnected parts of each classifying space.
\\
}

\begin{tabular}{lcl}
(a) complex classes & &
(b) real classes
\\
\begin{tabular}[t]{cccc}
\hline \hline
class & chiral & $C_q$ & $\pi_0(C_q)$ \\
\hline 
A &	0 & $C_0$ & $\mathbb{Z}$ \\  
AIII &	1 & $C_1$ & 0 \\  
\hline \hline
\end{tabular}
& ~ &
\begin{tabular}[t]{ccccc}
\hline \hline
class & TRS & PHS & $R_q$ & $\pi_0(R_q)$ \\
\hline
AI   & +1 & 0  & $R_0$ & $\mathbb{Z}$ \\  
BDI  & +1 & +1 & $R_1$ & $\mathbb{Z}_2$ \\  
D    & 0  & +1 & $R_2$ & $\mathbb{Z}_2$ \\  
DIII & $-1$ & +1 & $R_3$ & 0 \\  
AII  & $-1$ & 0  & $R_4$ & $\mathbb{Z}$ \\  
CII  & $-1$ & $-1$ & $R_5$ & 0 \\  
C    & 0  & $-1$ & $R_6$ & 0 \\  
CI   & +1 & $-1$ & $R_7$ & 0 \\  
\hline \hline
\end{tabular}
\end{tabular}
\label{tab-bott}
\end{center}
\end{table}

\subsection{Clifford algebras and their extentions}
We are going to classify topological insulators and superconductors
with additional reflection symmetries using Clifford algebras,
which are algebras with generators which anti-commute
with each other.

A complex Clifford algebra $Cl_n$ has
$n$ generators $e_i$ satisfying 
\begin{equation}
\{e_i,e_j\}=2\delta_{i,j}.
\label{eq:e_i,e_j complex}
\end{equation}
Linear combination of their products
$e_1^{p_1}e_2^{p_2}\cdots e_n^{p_n}$ ($p_i=0, 1$) multiplied 
by complex numbers form a $2^n$-dimensional complex vector space,
accompanied with the multiplication law (\ref{eq:e_i,e_j complex}).

A real Clifford algebra $Cl_{p,q}$ 
has $p+q$ generators satisfying 
\begin{align}
& \{e_i,e_j\}=0 ~~~ (i\neq j), \n
& e_i^2=
\begin{cases}
-1, & 1 \le i \le  p, \\
+1, & p+1 \le i \le  p+q. \\
\end{cases}
\label{eq:e_i,e_j real}
\end{align}
Their products with real coefficients form a $2^{p+q}$-dimensional
real vector space, accompanied with the multiplication
law (\ref{eq:e_i,e_j real}).

Having introduced real and complex Clifford algebras,
we now explain our strategy for classifying topological insulators and
superconductors with additional symmetries, which is a natural extension of
Kitaev's approach.\cite{kitaev09}
Namely, we reduce the classification problem of gapped free fermion systems
to that of possible extensions of
Clifford algebras which are generated by discrete symmetry operators
and a Hamiltonian with flattened spectra.

We begin with topological classification of zero-dimensional
systems (i.e., systems confined in a finite region) with some symmetries.
We first perform ``spectral flattening" of Hamiltonian $H$ with an energy gap,
i.e., we continuously deform eigenenergies above the gap to $+1$
and those below the gap to $-1$ while preserving wavefunctions.
This is a continuous deformation of the Hamiltonian and
does not change its topological properties.
Next we express symmetry constraints as generators $\{e_i\}$
of a Clifford algebra.
The relevant Clifford algebras for the complex and real AZ symmetry classes
(Table \ref{tab-bott}) are complex and real Clifford algebras, respectively.
We consider a matrix representation 
(of sufficiently large dimension)
of the Clifford algebra.
We then consider extending the algebra
by adding a generator $e_0$ which is obtained from the flattened Hamiltonian.
That is, for a fixed representation of symmetry constraints $\{e_i\}$, 
we look for possible representations of a new additional generator
$e_0$ .
The set of these representations for $e_0$ forms
a ``classifying space",
denoted as $C_q$ and $R_q$ for complex and real symmetry classes, respectively
(see Table \ref{tab-clifford} in Appendix \ref{appendix-Cl}).
Now, topologically distinct states correspond to topologically distinct
extensions of the algebra, and classification of them can be found from
a zero-th homotopy group of a classifying space $\pi_0(C_q)$ or $\pi_0(R_q)$,
i.e., the number of disconnected parts of $C_q$ or $R_q$.
The resulting classification for the 10 AZ classes
is summarized in Table~\ref{tab-bott},
whose explicit constructions are given in Sec~\ref{sec:AZ}.
We will apply this program to complete topological classification
in the presence of additional reflection symmetries
in the following sections.
Once the zero-dimensional systems are classified,
the classification of $d$-dimensional systems can be achieved
by considering $\pi_0(C_{q-d})$ and $\pi_0(R_{q-d})$,
as shown by Kitaev using K-theory.\cite{kitaev09}
In Sec.~\ref{sec:dimensional shift} we will give an alternative explanation
of the dimensional shift ($q\to q-d$)
for massive Dirac Hamiltonians in $d$ dimensions.

\subsection{Examples of classifying spaces}
In order to gain intuitions about classifying spaces and topological
invariants, let us look at a couple of examples of massive Dirac Hamiltonians
and discuss their classifying spaces by considering
what kind of Dirac mass term $\gamma_0$
is allowed in specific models.
The classifying space corresponds to a set of allowed Dirac mass terms.

First we consider a two-dimensional system in class A,
which is described by a $2N$ by $2N$ Hamiltonian
\begin{align}
H_{\t{2D}}= k_x \sigma_x \tensor 1_N + k_y \sigma_y \tensor 1_N + \gamma_0,
\end{align}
with momenta $k_i$, 2 by 2 Pauli matrices $\sigma_i$, 
a $N$ by $N$ identity matrix $1_N$,
and a mass term $\gamma_0$.
The classifying space corresponds to a set of allowed
Dirac mass term $\gamma_0$.
Since $\gamma_0$ anti-commutes with the kinetic terms,
$\gamma_0$ should have the form 
\begin{align}
\gamma_0=\sigma_z \tensor A,
\end{align}
where $A$ is a $N$ by $N$ Hermitian matrix and is normalized such that
it squares to $1_N$.
We can diagonalize $A$ with a unitary matrix $U$ as
\begin{align}
A=
U
I_{n,m}
U^\dagger , ~~~
I_{n,m}=
\begin{pmatrix}
1_n & 0 \\
0 & -1_m \\
\end{pmatrix},
\end{align}
with $N=n+m$.
$I_{n,m}$ is a diagonal matrix 
whose diagonal entries consist of $+1$ and $-1$,
appearing $n$ and $m$ times for each.
For a given $n$, 
$A$ is determined by a choice of $U$ from a unitary group $U(N)$,
but there is a redundancy 
in the choice of bases in each eigenspace of $\pm 1$.
Thus the set of $A$ with fixed $n$ and $m$ corresponds to
a complex Grassmanian $U(n+m)/U(n) \times U(m)$.
The total classifying space is a union of $U(n+m)/U(n) \times U(m)$
of different values of $n$.
If we assume $N$ to be sufficiently large, 
complex Grassmanians each labeled with $n$ become almost the same,
and
we can write the total classifying space as
\begin{align}
(U(n+m)/U(n) \times U(m)) \times \mathbb{Z},
\end{align}
which is exactly $C_0$ in Table \ref{tab-clifford}.
Since each complex Grassmanian is a connected manifold, 
each disconnected part of the classifying space 
(topologically distinct states) is specified by $n$,
the number of eigenvalues $+1$ of $A$.
Actually, $n$ coincides with the Chern number defined for the
2-dimensional system with a proper regularization.
We regularize the Dirac Hamiltonian by adding a $k^2$ term as
$H_{\t{2D}}-C k^2 \sigma_z \tensor 1_N$ with a small positive coefficient $C$. 
If we assume $A=I_{n,m}$ for simplicity,
then the Hamiltonian decouples into $N$ copies as
$H_{2D}=\oplus_{i=1}^N H_i$, where
\begin{align}
H_i=k_x \sigma_x + k_y \sigma_y + (\epsilon_i -C k^2) \sigma_z,
\label{eq:2d-classA}
\end{align}
with $\epsilon_i=+1$ for $1\le i \le n$, 
$\epsilon_i=-1$ for $n+1\le i \le N$.
Now the Chern number for $H_i$ is +1 for 
$\epsilon_i=+1$ and 0 for $\epsilon_i=-1$.
Therefore the total Chern number is $n$.

Next let us consider a one-dimensional system in class A,
with 
a $2N$ by $2N$ Hamiltonian
\begin{align}
H_{\t{1D}}= k_x \sigma_z \tensor 1_N + \gamma_0.
\end{align}
The Dirac mass term $\gamma_0$ should be Hermitian,
 anti-commute with $\sigma_z$,
and square to $1_{2N}$,
which requires $\gamma_0$ to be written as
\begin{align}
\gamma_0=
\begin{pmatrix}
0 & U \\
U^\dagger & 0 \\
\end{pmatrix},
\end{align}
where $U$ is a $N$ by $N$ unitary matrix.
Thus each choice of $U$ specifies the mass term $\gamma_0$ 
and the classifying space is given by $U(N)$,
which is $C_1$ in Table \ref{tab-clifford}.
Since the unitary group $U(N)$ is connected,
we can 
deform one state into another without closing the energy gap;
there exists only one phase which is topologically trivial.
This trivial classification is a consequence of 
the existence of
two mass terms that anti-commute with each other 
in the minimal Dirac model.
Let us consider a minimal 2 by 2 Hamiltonian,
\begin{align}
H_{\t{1D}}= k_x \sigma_z + m_1 \sigma_x + m_2 \sigma_y.
\end{align}
When there is only one allowed mass term ($m_0 \sigma_z$)
as in the previous example [Eq.\ (\ref{eq:2d-classA})],
two states with different signs of the unique mass term
are topologically distinct,
and 
changing the sign of $m_0$ is only possible via a point $m_0=0$
where the bulk gap closes.
On the other hand, if we have two masses ($m_1$ and $m_2$),
the two states with masses $(m_1,m_2)=(\pm 1,0)$ are connected 
through a rotation in the plane of $(m_1,m_2)$,
i.e., we can deform one to the other without closing the bulk gap.
This deformation is regarded as a rotation in $U(N)$ with $N=1$.

\section{Classification for AZ classes\label{sec:AZ}}
In this section, we give a concise review of the topological classification
of the ten AZ symmetry classes in zero spatial dimension
in terms of an extension problem of the Clifford algebra,
in a way complementary to the original Kitaev's paper.\cite{kitaev09}
The two complex classes are classified with complex Clifford algebras
while the eight real classes are classified with real Clifford algebras.
This section will serve as a starting point for the topological
classification in the presence of additional reflection symmetries
in the following sections.

\subsection{Complex classes}
We start with classification of the complex AZ classes (A and AIII)
in terms of complex Clifford algebras.
The extension problem for complex Clifford algebra $Cl_n \to Cl_{n+1}$ is
characterized by a classifying space $C_n$.\cite{kitaev09,karoubi}

When a zero-dimensional system is a member of class AIII,
its Hamiltonian $H$ satisfies the chiral symmetry relation
\begin{equation}
\{H,\Gamma\}=0,
\label{eq:chiral}
\end{equation}
where $\Gamma$ is a unitary operator.
After spectral flattening, the zero-dimensional Hamiltonian $H$ has
eigenvalues $\pm1$, and we set $e_0:=H$.
We note that the relation (\ref{eq:chiral}) is not affected by
the spectral flattening.
Without loss of generality we may assume $\Gamma^2=1$ and
regard $e_1:=\Gamma$ as a generator of complex Clifford algebra $Cl_1$.
We then consider extending the complex Clifford algebra
$Cl_1$ to $Cl_2$ by adding the generator $e_0$ to the algebra $Cl_1$
(Table~\ref{tab-complex-AZ}).
A set of the possible representations of $e_0$ in the extended algebras
form the classifying space $C_1$.
Since $\pi_0(C_1)=0$, zero-dimensional systems in class AIII are
topologically trivial (Table~\ref{tab-bott}).

For Hamiltonians in class A, we begin with complex Clifford
algebra $Cl_0$.
We consider the extension of
$Cl_0$ to $Cl_1$ with $e_0=H$, whose possible representations
form the classifying space $C_0$ (Table~\ref{tab-complex-AZ}).
We then find from $\pi_0(C_0)=\mathbb{Z}$ that zero-dimensional systems
in class A are characterized by an integer topological index
(Table~\ref{tab-bott}).


\begin{table}[tb]
\caption{Classification of the complex AZ classes (A and AIII)
in zero dimension
from the extension of complex Clifford algebras.
The last column shows classifying spaces $C_q$.
}
\begin{center}
\begin{tabular}{cccc}
\hline\hline
class & generators & extension & classifying space \\
\hline
A  & $e_0=H$ & $Cl_0 \to Cl_1$ & $C_0$ \\
AIII ~ & $e_0=H,~e_1=\Gamma$ ~ & $Cl_1 \to Cl_2$  & $C_1$ \\
\hline\hline
\end{tabular}
\end{center}
\label{tab-complex-AZ}
\end{table}

\subsection{Real classes}
Next we review classification of the real AZ classes
in zero spatial dimension
in terms of real Clifford algebras.
Time-reversal symmetry (TRS) and particle-hole symmetry (PHS)
of Hamiltonian $H$ are written as
\begin{align}
T^{-1}H T &=H, & C^{-1}H C &=-H,
\label{eq:TRS and PHS}
\end{align}
with anti-unitary operators $T$ and $C$, respectively.
These relations are not affected by spectral flattening of $H$,
\begin{equation}
H^2=1.
\label{eq:H2=1}
\end{equation}
Without loss of generality
we can assume
\begin{equation}
[T,C]=0,
\label{eq:[T,C]=0}
\end{equation}
and we have
\begin{equation}
T^2=\epsilon_T,
\qquad
C^2=\epsilon_C,
\label{eq:T2 and C2}
\end{equation}
where $\epsilon_T$ and $\epsilon_C$ are either $+1$ or $-1$;
see Table~\ref{tab-real-AZ}.
Since both $T$ and $C$ involve complex conjugation $\mathcal{K}$,
we introduce an operator $J$ representing the imaginary unit ``$\, i\,$"
so that we can treat complex structure algebraically
in real Clifford algebras.
We thus impose the operator $J$ to satisfy the relations
\begin{equation}
J^2=-1,\qquad
\{T,J\}=\{C,J\}=[H,J]=0,
\label{eq:J}
\end{equation}
as expected for ``$\, i\,$''.


\begin{table}[tb]
\caption{
Classification for the real AZ classes in zero dimension
from the extension of real Clifford algebras.
The second column $(\e_T,\e_C)$ shows the sign of squared symmetry
operators $(T^2,C^2)$,
where the absence of the symmetry is denoted by ``$0$".
}
\begin{center}
\begin{tabular}{cccc}
\hline\hline
class & ~ $(\e_T,\e_C)$ ~ & extension & ~ classifying space \\
\hline
AI  & $(+,0)$ & $Cl_{0,2} \to Cl_{1,2}$ & $R_0$ \\
AII & $(-,0)$ & $Cl_{2,0} \to Cl_{3,0}$ & $R_4$ \\
\hline
D & $(0,+)$ & $Cl_{0,2} \to Cl_{0,3}$ & $R_2$ \\
C & $(0,-)$ & $Cl_{2,0} \to Cl_{2,1}$ & $R_{-2}\simeq R_6$ \\
\hline
BDI & $(+,+)$ & $Cl_{1,2} \to Cl_{1,3}$ & $R_1$ \\
DIII & $(-,+)$ & $Cl_{0,3} \to Cl_{0,4}$ & $R_3$ \\
CII & $(-,-)$ & $Cl_{3,0} \to Cl_{3,1}$ & $R_{-3}\simeq R_5$ \\
CI & $(+,-)$ & $Cl_{2,1} \to Cl_{2,2}$ & $R_{-1}\simeq R_7$ \\
\hline\hline
\end{tabular}
\end{center}
\label{tab-real-AZ}
\end{table}

Equations (\ref{eq:TRS and PHS})--(\ref{eq:J}) are used
to define real Clifford algebras $Cl_{p,q}$.
According to the absence or presence of the TRS and PHS,
we have a different set of generators for real Clifford algebra
in each class:
\begin{enumerate}
\item[(i)] $T$ only (AI and AII): $\{e_1,e_2\}\to\{e_0,e_1,e_2\}$,
where%
\begin{subequations}
\label{generators-Tonly}
\begin{equation}
e_0=JH, \qquad e_1=T, \qquad e_2=TJ,
\end{equation}
with
\begin{equation}
e_0^2=-1, \qquad e_1^2=\e_T, \qquad e_2^2=\e_T.  
\end{equation}
\end{subequations}
\item[(ii)] $C$ only (C and D): $\{e_1,e_2\}\to\{e_0,e_1,e_2\}$,
where
\begin{subequations}
\label{generators-Conly}
\begin{equation}
e_0=H, \qquad e_1=C, \qquad e_2=CJ,
\end{equation}
with
\begin{equation}
e_0^2=1, \qquad e_1^2=\e_C, \qquad e_2^2=\e_C. 
\end{equation}
\end{subequations}
\item[(iii)] Both $T$ and $C$ (BDI, DIII, CII, and CI):\\
$\{e_1,e_2,e_3\}\to\{e_0,e_1,e_2,e_3\}$, where
\begin{subequations}
\label{generators-TandC}
\begin{equation}
e_0=H, \quad e_1=C, \quad e_2=CJ, \quad e_3=TCJ,
\end{equation}
with
\begin{equation}
e_0^2=1, \quad e_1^2=\e_C, \quad e_2^2=\e_C, \quad e_3^2=-\e_T\e_C.
\end{equation}
\end{subequations}
\end{enumerate}
Before adding the generator $e_0$, we have real Clifford algebras
$Cl_{2,0}$ or $Cl_{0,2}$ for the cases (i) and (ii), and $Cl_{1,2}$,
$Cl_{2,1}$, $Cl_{0,3}$, or $Cl_{3,0}$ for the case (iii), depending
on the sign of $\epsilon_T$ and $\epsilon_C$ (Table~\ref{tab-real-AZ}).
We then consider extension of the real Clifford algebras by adding $e_0$.
We shall distinguish two cases, $e_0^2=+1$ and $e_0^2=-1$.
The classifying space for the extension
$Cl_{p,q} \to Cl_{p,q+1}$ ($e_0^2=+1$) is known to
be given by $R_{q-p}$,
while that for $Cl_{p,q} \to Cl_{p+1,q}$ ($e_0^2=-1$)
is given by $R_{p+2-q}$. \cite{kitaev09,karoubi}
The latter can be understood by noting that
we have an isomorphism $Cl_{p,q} \tensor \mathbb{R}(2) \simeq Cl_{q,p+2}$,
where $\mathbb{R}(2)$ is an algebra of 2 by 2 real matrices (Appendix \ref{appendix-Cl}).
By taking tensor product with $\mathbb{R}(2)$
(which does not affect the extension problem),
the extension $Cl_{p,q}\to Cl_{p+1,q}$
is mapped to the extension $Cl_{q,p+2} \to Cl_{q,p+3}$,
whose classifying space is $R_{p+2-q}$.
From the Bott periodicity,\cite{kitaev09,karoubi}
the classifying space has a periodic structure $R_q \simeq R_{q+8}$,
and the eight real symmetry classes fall into eight distinct
classifying spaces, as listed in Table~\ref{tab-real-AZ}
(see Appendix~\ref{appendix-Cl}).
Finally we find topological classification of each AZ class
from zeroth homotopy group of the classifying spaces $\pi_0(R_q)$
(Table~\ref{tab-bott}).

\section{additional symmetry\label{sec:additional}}
In this section, we study how the topological properties change
when an additional symmetry is imposed on each symmetry class.
Here we concentrate on an additional symmetry
denoted by a unitary operator $M$ that
anti-commutes with Hamiltonian $H$,
\begin{equation}
\{H,M\}=0.
\label{eq:[H,M]=0}
\end{equation}
As we will see later in the following sections,
this situation has several interesting applications.
The additional symmetry is represented as a new generator in
Clifford algebras, leading to modification of the Clifford algebras and
topological classification.
While we consider zero-dimensional Hamiltonian $H$ in this section,
we note that,
once we find the classifying space for zero dimension as $C_q$ or $R_q$,
the topological classification for $d$ dimensions is given by 
$\pi_0(C_{q-d})$ or $\pi_0(R_{q-d})$,
according to K-theory.\cite{kitaev09}

\subsection{Complex classes}

\begin{table}[tb]
\caption{Shifts of classifying spaces due to an additional symmetry
for complex classes in zero spatial dimension.}
\begin{tabular}{cccc}
\hline\hline
class & $\eta_\Gamma$ & relations & shift of $C_n$ \\
\hline
A & & $\{e_0,M\}=0$ & $+1$ \\
AIII & ~ $+$ ~ & $[e_0,\Gamma M]=[e_1,\Gamma M]=0$ & $0$ \\
AIII &  $-$ & $\{e_0,M\}=\{e_1,M\}=0$ & $+1$ \\
\hline\hline
\end{tabular}
\label{tab-M-complex}
\end{table}

We first consider zero-dimensional systems in complex classes.

For systems originally in class A, $M$ serves as a chiral symmetry operator.
Hence the relevant classifying space for class A
changes from $C_0$ to $C_1$ upon addition of the symmetry $M$.

For systems in class AIII with chiral symmetry $\Gamma$,
we impose the condition
\begin{equation}
\Gamma M=\eta_\Gamma M \Gamma,
\end{equation}
where the signature $\eta_\Gamma$ designates commutation $(+)$ or
anti-commutation $(-)$ relation between $M$ and $\Gamma$.
If $\eta_\Gamma=-1$, then $M$ serves as an additional generator
to the complex Clifford algebra, and we need to consider an extension
problem $Cl_2\to Cl_3$, instead of the original $Cl_1\to Cl_2$.
Hence the classifying space is shifted by 1, from $C_1$ to $C_2\simeq C_0$.
Here we have used the Bott periodicity,\cite{kitaev09,karoubi}
$C_n \simeq C_{n+2}$.

On the other hand, when $\eta_\Gamma=+1$,
the product $\Gamma M$ commutes with the original generators $\Gamma$ and $H$.
Then, in each eigenspace of $\Gamma M$, we have the same extension problem
of a complex Clifford algebra as before ($Cl_1\to Cl_2$),
and the topological classification is not changed from Table~\ref{tab-bott}.

\subsection{Real classes}
Next we consider zero-dimensional systems in the eight real classes
with an additional symmetry $M$.
We require $M$ to satisfy
\begin{equation}
[J,M]=0,
\qquad
M^2=1.
\end{equation}
Furthermore, for systems which are invariant under time-reversal or
particle-hole transformation,
we assume
\begin{align}
TM=\eta_T MT, \qquad
CM=\eta_C MC,
\label{eq:TMandCM}
\end{align}
where $\eta_T$ and $\eta_C$ designate commutation ($+$) or
anti-commutation ($-$) relations with $T$ and $C$.


\begin{table}[tb]
\caption{Classification of real symmetry classes in the presence
of an additional symmetry $M$ in zero dimension.
$(\eta_T,\eta_C)$ dictates commutation $(+)$ or anti-commutation $(-)$
relation with $T$ and $C$ operators [Eq.\ (\ref{eq:TMandCM})],
or absence of symmetry denoted by ``$0$".
We distinguish the following two cases:
(a) The additional symmetry operator $M$ provides a new generator ($\tilde e$)
and causes a shift in the classifying space;
(b) The additional symmetry operator $M$ provides an operator ($\widetilde M$)
commuting with all pre-existing generators.
}
(a)\\
\begin{tabular}{ccccc}
\hline\hline
class & $(\eta_T,\eta_C)$ & $\tilde e$ & $\tilde{e}^2$ & ~ shift of $R_q$ \\
\hline
\multirow{2}{*}{AI, AII}
&$(+,0)$ & $JM$ & $-1$ & $+1$ \\
&$(-,0)$ & $M$ & $+1$ & $-1$ \\
\hline
\multirow{2}{*}{D, C}
&$(0,+)$ & $JM$ & $-1$ & $-1$ \\
&$(0,-)$ & $M$ & $+1$ & $+1$ \\
\hline
\multirow{2}{*}{BDI, DIII, CII, CI}
&$(+,-)$ & $M$ & $+1$ & $+1$ \\
&$(-,+)$ & ~ $JM$ ~ & $-1$ & $-1$ \\
\hline\hline
\end{tabular}
\\ \vspace{1em}

(b)\\
\begin{tabular}{ccccc}
\hline\hline
class & $(\eta_T,\eta_C)$ & $\widetilde M$ & $\widetilde M^2$ &
classifying space \\
\hline
\multirow{2}{*}{BDI, CII}
&$(+,+)$ & $TCM$ & $+1$ & ~ same as in Table~\ref{tab-real-AZ} \\
&$(-,-)$ & $TCJM$ & $-1$ & $C_1$ \\
\hline
\multirow{2}{*}{DIII, CI}
&$(+,+)$ & $TCM$ & $-1$ & $C_1$ \\
&$(-,-)$ & ~ $TCJM$ ~ & $+1$ & ~ same as in Table~\ref{tab-real-AZ} \\
\hline\hline
\end{tabular}
\label{tab-M-real}
\end{table}

If the system has either one of TRS and PHS,
or if it has both TRS and PHS with $(\eta_T,\eta_C)=(+,-)$ or $(-,+)$,
then we can construct an additional generator $\tilde e$
from the symmetry $M$ given in Table~\ref{tab-M-real}(a),
which should be added to the set of generators listed in
Eqs.~(\ref{generators-Tonly})--(\ref{generators-TandC}).
By considering the extension problem of Clifford algebras,
$\{e_1,\ldots,\tilde e\}\to\{e_0,e_1,\ldots,\tilde e\}$, we find that
the index $q$ of the classifying space $R_q$ should be shifted
by $\pm 1$, according to the sign of $\tilde e^2$,
as listed in Table~\ref{tab-M-real}(a).

On the other hand,
if the system has both TRS and PHS with $(\eta_T,\eta_C)=(+,+)$ or $(-,-)$,
then we can construct,
not an additional generator for Clifford algebras,
but an operator $\widetilde M$ which commutes with
all generators of the relevant Clifford algebras;
see Table~\ref{tab-M-real}(b).
We have the following two cases according to the sign of ${\widetilde M}^2$.

(1) $\widetilde M^2=+1$: 
The Hilbert space splits into two eigenspaces of $\widetilde M$.
For each eigenspace of $\widetilde M$,
we have the same Clifford algebra and classifying space as
in Table~\ref{tab-real-AZ}, and the topological classification is not changed
from Table~\ref{tab-bott}.

(2) $\widetilde M^2=-1$: $\widetilde M$ effectively behaves like ``$\, i\,$"
and introduces a complex structure into the real algebra. 
This situation is formally written as 
$Cl_{p,q}\tensor_{\mathbb{R}} Cl_{1,0} \simeq Cl_{p,q}\tensor_{\mathbb{R}}\mathbb{C}
\simeq Cl_{p+q}$.
Then the classification is given by an extension problem
of a complex Clifford algebra $Cl_3 \to Cl_4$,
which falls into a classifying space $C_3 \simeq C_1$.
This means that the additional symmetry has changed real symmetry
classes into class AIII.

Changes in topological classification of real classes
induced by an additional symmetry in zero spatial dimension
are summarized in Table~\ref{tab-M-real}.

\subsection{Commuting symmetries $[H,M]=0$}
So far we have studied additional symmetries that anti-commute
with Hamiltonian. 
Here we briefly discuss additional symmetries that satisfy
\begin{align}
[H,M]=0, \qquad M^2=1.
\end{align}
For example, this situation happens in 1D systems with mirror lines,
or in 2D systems with mirror planes as discussed
in Ref.~\onlinecite{zhang-kane-tmsc,ueno2013symmetry}.
We show below that, upon block diagonalization of $H$ with respect to $M$,
symmetry classes can change 
if actions of the generic symmetries are not closed in eigenspaces of $M$.

We start with complex classes.
In class A, the block diagonalization does not change the symmetry class
nor the classification.
In class AIII, when the chiral symmetry also commutes as $[\Gamma, M]=0$,
the classification does not change.
On the other hand, when the chiral symmetry anti-commutes as $\{\Gamma, M\}=0$,
the chiral symmetry does not hold in subblocks and the symmetry class
changes into class A.

Let us move on to real classes.
(i) $T$ only (AI and AII) or $C$ only (C and D):
When the additional symmetry $M$ commutes with the generic symmetry $T$ or $C$,
the block diagonalization does not change the symmetry class and,
hence, the classification.
The situation changes for the additional symmetry $M$ that anti-commutes
with the generic symmetry. Since TRS or PHS does not hold
in each eigenspace of $M$, the symmetry class changes into class A.
In terms of Clifford algebras, this situation is written as an extension
problem $\{e_1,e_2\}\tensor{JM} \to \{e_0,e_1,e_2\}\tensor{JM}$,
where we adopt a new generator $JM$ that commutes with $T$ or $C$
but squares to $-1$. 
Thus the extension
$Cl_{p,q}\tensor \mathbb{C} \to Cl_{p,q+1}\tensor \mathbb{C}$
turns into that of complex algebras $Cl_{p+q} \to Cl_{p+q+1}$.
Then the relevant classifying space is a complex one with
$C_{p+q} \simeq C_0$ from Table \ref{tab-real-AZ},
which indicates that the symmetry class changes into class A.

(ii) Both $T$ and $C$ (BDI, DIII, CII, and CI):
Since we have a chiral symmetry $\Gamma$ in this case,
we can construct an operator that anti-commutes with the Hamiltonian
and squares to $+1$ (either $M'=\Gamma M$ or $M'=J\Gamma M$),
with which we can use the result of Sec.~\ref{sec:additional}B.
We will later discuss an example for this type of symmetry
in Sec.~\ref{sec:examples}B.

\section{Dirac model and dimensional shift\label{sec:dimensional shift}}
As Kitaev has shown using K-theory,\cite{kitaev09,karoubi}
once classifying spaces, $C_q$ or $R_q$, is understood for each
AZ symmetry class in zero dimension,
topological classification of $d$-dimensional systems ($d\ge1$)
of that class is directly found from
$\pi_0(C_{q-d})$ or $\pi_0(R_{q-d})$.
In this section we consider massive Dirac Hamiltonians to
see this dimensional shift explicitly in terms of Clifford algebras 
with additional symmetries;
similar formulations are given in
Ref.~\onlinecite{stone2011symmetries,wen-SPT-fermion12,abramovici-kalugin}.

Let us consider a massive Dirac Hamiltonian in $d$ dimensions,
\begin{equation}
H_d=\sum_{i=1}^d \gamma_i k_i +H,
\end{equation}
where $k_i$ is a momentum in the $i$th direction
and $H$ is a Dirac mass term which should satisfy appropriate
symmetry relations (\ref{eq:chiral}) or (\ref{eq:TRS and PHS})
for each symmetry class.
We can impose the constraint $H^2=1$
(by taking the Dirac mass as a unit of energy).
The gamma matrices obey the anticommutation relations
\begin{equation}
\{\gamma_i,\gamma_j\}=2\delta_{i,j},\qquad
\{H,\gamma_i\}=0.
\end{equation}
Furthermore, the TRS and PHS (if present) of the Hamiltonian $H_d$ imply
\begin{equation}
\{T,\gamma_i\}=0,
\qquad
[C,\gamma_i]=0,
\label{eq:TC vs gamma_i}
\end{equation}
because complex conjugation $(\mathcal{K})$ changes $k_i$ to $-k_i$.
The chiral symmetry (if present) imposes
\begin{equation}
\{\Gamma,\gamma_i\}=0.
\end{equation}

Now we show that topological classification of the Dirac Hamiltonian $H_d$
is obtained from $\pi_0(C_{q-d})$ or $\pi_0(R_{q-d})$ when we study the
extension (by adding $e_0:=H$) of Clifford algebras
generated by generic symmetry operators and the $\gamma_i$'s.
We proceed by induction.
Suppose that the classifying space for the Dirac mass term $H$ 
in the $(d-1)$-dimensional system
$H_{d-1}$ is found to be $C_{q-d+1}$ or $R_{q-d+1}$.
The $d$-dimensional Hamiltonian $H_d$ has an additional operator $M=\gamma_d$,
which can be considered as an additional symmetry constraint discussed
in Sec.~\ref{sec:additional}.
For real classes, when the Dirac mass term $H$ in $H_d$ is invariant
under $T$ and/or $C$,
the additional operator $M$ anti-commuting with $H$
 has the signature $\eta_T=-1$ and/or $\eta_C=+1$ from
Eq.~(\ref{eq:TC vs gamma_i}).
We find from Table~\ref{tab-M-real}(a) that the addition of $M$ induces
a shift of the relevant classifying space by $-1$; we have $R_{q-d}$.
For complex classes A and AIII ($\eta_\Gamma=-1$),
we find from Table~\ref{tab-M-complex} that
the addition of $M$ induces a shift by $+1=-1 \t{ (mod 2)}$
in complex classifying spaces; we have $C_{q-d}$.

\section{Reflection symmetry \label{sec:reflection}}
Let us discuss topological classification in the presence of reflection
symmetry using the Dirac model studied in the preceding section.
Our approach is similar but complementary to the one
in Ref.~\onlinecite{chiu13}, where topological invariants are
explicitly constructed for Dirac models.
We apply the classification theory in terms of Clifford algebras,
considering a reflection symmetry as a special case of
additional symmetries discussed in Sec.~\ref{sec:additional}.
For the sake of simplicity, we first assume translation symmetry
and exclude terms which depend on spatial coordinates
(such as mass terms of CDW type)
in the Dirac Hamiltonian.
We will comment on effects of relaxing this condition at the
end of this section and in Appendixes~\ref{App-reflection}
and \ref{App-stability}.

Let us assume that the Dirac Hamiltonian $H_d$ is invariant under
reflection $R$ in the $l$th direction,
where momenta $k_j$'s are changed as $k_j\to(-1)^{\delta_{j,l}}k_j$.
It follows from $[R,H_d]=0$ that
$\{\gamma_l,R\}=0$ and $[\gamma_j, R]=0$ for $j\ne l$.
We can always set $R^2=1$.
Then, as an additional symmetry operator $M$,
we can take
\begin{equation}
M=J\gamma_l R,
\label{eq:M=igamma R}
\end{equation}
which satisfy $M^2=1$, $\{H,M\}=0$, and $\{\gamma_j,M\}=0$ ($1\le j\le d$),
where again $H$ is a Dirac mass term satisfying symmetry relation
(\ref{eq:chiral}) or (\ref{eq:TRS and PHS}) appropriate for each
symmetry class.

Note that the commutation relations of $R$ with $C$ and $\Gamma$
are different from those of $M$.
Namely, for real classes, when $R$ obeys the relations
$RT=\tilde \eta_T TR$ and/or $RC=\tilde \eta_C CR$
with $\tilde\eta_{T,C}=+$ or $-$,
$M$ has the signatures
$(\eta_T,\eta_C)=(\tilde\eta_T,-\tilde\eta_C)$.
Similarly, for class AIII,
we have $\eta_\Gamma=-\tilde \eta_\Gamma$,
where $\tilde\eta_\Gamma=+$ or $-$ is specified by
the relation $R\Gamma=\tilde \eta_\Gamma \Gamma R$.
Now that we have defined $M$ with the signatures $(\eta_T,\eta_C,\eta_\Gamma)$,
we can use Tables~\ref{tab-M-complex} and \ref{tab-M-real} and
the dimensional shift discussed in Sec.~\ref{sec:dimensional shift},
to obtain topological classification in the presence of
the reflection symmetry $R$.

Let us discuss in more detail the consequence of reflection symmetry
for each class.
We begin with complex classes.
First, class A turns into AIII with an effective chiral symmetry $M$.
For class AIII, if we denote $R$ with its commutation relation
with $\Gamma$ as $R^{\tilde \eta_\Gamma}$, we find that $R^+$ changes
the classifying space to $C_0$,
while  $R^-$ does not change the classification.
Now let us move on to real symmetry classes.
We write the $R$ operator with a superscript showing its
commutation relations with $T$ or $C$ as
$R^{\tilde \eta_T}, R^{\tilde \eta_C},R^{\tilde \eta_T \tilde \eta_C}$
for symmetry classes with $T$ only, $C$ only, and both $T$ and $C$,
respectively.
Suppose that the original classifying space for a given real symmetry
class is $R_{q-d}$ for $d$ dimensions.
We find from Table~\ref{tab-M-real} that
$R^{+}$ and $R^{++}$ shift the classifying space by $+1$ to $R_{q-d+1}$.
Thus we may say that $R^{+}$ and $R^{++}$ have the effect of decreasing
the spatial dimension by $1$.
In a similar way, $R^{-}$ and $R^{--}$ shift the classifying space by $-1$
to $R_{q-d-1}$ and effectively increase the spatial dimension by $1$.
As for $R^{+-}$,
topological classifications for BDI and CII
remain the same, while DIII and CI change into complex class AIII.
On the contrary, with $R^{-+}$, DIII and CI remain the same,
while BDI and CII change into complex class AIII.

\begin{table*}[tb]
\begin{center}
\caption{
Classification in the presence of a reflection symmetry.
The first column shows types of reflection symmetry,
where superscripts of $R$ show its commutation relations
with basic symmetry operators such as $\Gamma$, $T$, and $C$,
i.e.,
$R^{\tilde \eta_\Gamma}$ for the complex classes,
and $R^{\tilde \eta_T}, R^{\tilde \eta_C},R^{\tilde \eta_T \tilde \eta_C}$
for real classes with $T$ only, $C$ only, and both $T$ and $C$,
respectively.
The second column shows symmetry classes in which a given Hamiltonian
is classified without reflection symmetry taken into account.
The third column shows classifying spaces for zero-dimensional
Hamiltonian in the presence of reflection symmetry,
and the following columns show topological classifications for
spatial dimensions $d=0,1,2,\ldots,7$ (mod 8).
Note that the classifying spaces in the third column
are shifted from those listed in Table \ref{tab-bott}.
The ``$\mathbb{Z}_2$'' phases appearing under the reflection
$R^-$ or $R^{--}$ turn into topologically trivial (0),
when spatially non-uniform perturbations are applied
(see the discussion at the end of Sec.~\ref{sec:reflection}
and Appendix \ref{App-reflection}).
\\
}

\begin{tabular}[t]{ccccccccccc}
\hline \hline
Reflection & Class & $C_q$ or $R_q$ & $d=0$ & $d=1$ & $d=2$ & $d=3$
 & $d=4$ & $d=5$ & $d=6$ & $d=7$ \\
\hline
$R$  & A      &$C_1$& $0$ & $\mathbb{Z}$ & $0$ & $\mathbb{Z}$ & $0$
 & $\mathbb{Z}$ & $0$ & $\mathbb{Z}$ \\
$R^+$& AIII   &$C_0$& $\mathbb{Z}$ & $0$ & $\mathbb{Z}$ & $0$ & $\mathbb{Z}$
 & $0$ & $\mathbb{Z}$ & $0$ \\
$R^-$& AIII   &$C_1$& $0$ & $\mathbb{Z}$ & $0$ & $\mathbb{Z}$ & $0$
 & $\mathbb{Z}$ & $0$ & $\mathbb{Z}$ \\
\hline 
\multirow{8}{*}{$R^+,R^{++}$}
 & AI   &$R_1$& $\mathbb{Z}_2$ & $\mathbb{Z}$ & $0$ & $0$
 & $0$ & $\mathbb{Z}$ & $0$ & $\mathbb{Z}_2$ \\
 & BDI  &$R_2$& $\mathbb{Z}_2$ & $\mathbb{Z}_2$ & $\mathbb{Z}$ & $0$
 & $0$ & $0$ & $\mathbb{Z}$ & $0$ \\
 & D    &$R_3$& $0$ & $\mathbb{Z}_2$ & $\mathbb{Z}_2$ & $\mathbb{Z}$
 & $0$ & $0$ & $0$ & $\mathbb{Z}$ \\
 & DIII &$R_4$& $\mathbb{Z}$ & $0$ & $\mathbb{Z}_2$ & $\mathbb{Z}_2$
 & $\mathbb{Z}$ & $0$ & $0$ & $0$ \\
 & AII  &$R_5$& $0$ & $\mathbb{Z}$ & $0$ & $\mathbb{Z}_2$
 & $\mathbb{Z}_2$ & $\mathbb{Z}$ & $0$ & $0$  \\
 & CII  &$R_6$& $0$ & $0$ & $\mathbb{Z}$ & $0$
 & $\mathbb{Z}_2$ & $\mathbb{Z}_2$ & $\mathbb{Z}$ & $0$ \\
 & C    &$R_7$& $0$ & $0$  & $0$ & $\mathbb{Z}$ & $0$
 & $\mathbb{Z}_2$ & $\mathbb{Z}_2$ & $\mathbb{Z}$ \\
 & CI   &$R_0$& $\mathbb{Z}$ & $0$ & $0$  & $0$ &
 $\mathbb{Z}$ & $0$ & $\mathbb{Z}_2$ & $\mathbb{Z}_2$ \\
\hline
\multirow{8}{*}{$R^-,R^{--}$}
 & AI   &$R_7$& $0$ & $0$  & $0$ & $\mathbb{Z}$
 & $0$ & ``$\mathbb{Z}_2$'' & $\mathbb{Z}_2$ & $\mathbb{Z}$ \\
 & BDI  &$R_0$& $\mathbb{Z}$ & $0$ & $0$  & $0$
 & $\mathbb{Z}$ & $0$ & ``$\mathbb{Z}_2$'' & $\mathbb{Z}_2$ \\
 & D    &$R_1$& $\mathbb{Z}_2$ & $\mathbb{Z}$ & $0$
 & $0$ & $0$ & $\mathbb{Z}$ & $0$ & ``$\mathbb{Z}_2$'' \\
 & DIII &$R_2$& ``$\mathbb{Z}_2$'' & $\mathbb{Z}_2$ & $\mathbb{Z}$ & $0$
 & $0$ & $0$ & $\mathbb{Z}$ & $0$ \\
 & AII  &$R_3$& $0$ & ``$\mathbb{Z}_2$'' & $\mathbb{Z}_2$ & $\mathbb{Z}$
 & $0$ & $0$ & $0$ & $\mathbb{Z}$ \\
 & CII  &$R_4$& $\mathbb{Z}$ & $0$ & ``$\mathbb{Z}_2$'' & $\mathbb{Z}_2$
 & $\mathbb{Z}$ & $0$ & $0$ & $0$ \\
 & C    &$R_5$& $0$ & $\mathbb{Z}$ & $0$ & ``$\mathbb{Z}_2$''
 & $\mathbb{Z}_2$ & $\mathbb{Z}$ & $0$ & $0$  \\
 & CI   &$R_6$& $0$ & $0$ & $\mathbb{Z}$ & $0$
 & ``$\mathbb{Z}_2$'' & $\mathbb{Z}_2$ & $\mathbb{Z}$ & $0$ \\
\hline
$R^{+-}$ & BDI  &$R_1$& $\mathbb{Z}_2$ & $\mathbb{Z}$ & $0$ & $0$
 & $0$ & $\mathbb{Z}$ & $0$ & $\mathbb{Z}_2$ \\
$R^{-+}$ & DIII &$R_3$& $0$ & $\mathbb{Z}_2$ & $\mathbb{Z}_2$ & $\mathbb{Z}$
 & $0$ & $0$ & $0$ & $\mathbb{Z}$ \\
$R^{+-}$ & CII  &$R_5$& $0$ & $\mathbb{Z}$ & $0$ & $\mathbb{Z}_2$
 & $\mathbb{Z}_2$ & $\mathbb{Z}$ & $0$ & $0$  \\
$R^{-+}$ & CI   &$R_7$& $0$ & $0$  & $0$ & $\mathbb{Z}$
 & $0$ & $\mathbb{Z}_2$ & $\mathbb{Z}_2$ & $\mathbb{Z}$ \\
\hline
$R^{-+}$ & BDI, CII  &$C_1$& $0$ & $\mathbb{Z}$ & $0$ & $\mathbb{Z}$
 & $0$ & $\mathbb{Z}$ & $0$ & $\mathbb{Z}$ \\
$R^{+-}$ & DIII, CI  &$C_1$& $0$ & $\mathbb{Z}$ & $0$ & $\mathbb{Z}$
 & $0$ & $\mathbb{Z}$ & $0$ & $\mathbb{Z}$ \\
\hline \hline
\end{tabular}
\label{tab-reflection}
\end{center}
\end{table*}

Table \ref{tab-reflection} summarizes
classification in the presence of various types of reflection symmetries
for each spatial dimension.
The periodic structures are evident.

A brief comment is in order on the relation between the topological
classification of Hamiltonian in the bulk and the presence of
gapless boundary states.
In topological mirror insulators/superconductors where the bulk Hamiltonian
has a nontrivial topological index only in the presence of a reflection
symmetry, the existence of gapless states on a boundary
depends on whether or not the boundary preserves the reflection
symmetry.\cite{fu-tci11}
When the presence of a boundary is compatible with reflection symmetry
(e.g., when the boundary is normal to a mirror plane), 
gapless boundary states are stable and protected by the above classification.
On the other hand, if a boundary breaks reflection symmetry,
the boundary states are generally gapped.
This is the case for one-dimensional systems
where the presence of an edge breaks the reflection symmetry
(unless the mirror plane is parallel to the one-dimensional system itself).

So far we have assumed the translation symmetry.
However, as pointed out in Ref.~\onlinecite{chiu13},
if the condition of translation symmetry is removed,
the ``second descendant''\cite{ryu-njp10} $\mathbb{Z}_2$ states
under the reflection symmetry $R^-$ or $R^{--}$
can be adiabatically deformed into a topologically trivial insulator
by introducing an extra mass term with a finite wave number.
These unstable $\mathbb{Z}_2$ phases are denoted by ``$\mathbb{Z}_2$'' in
Table \ref{tab-reflection}.
Once we replace ``$\mathbb{Z}_2$'' with 0,
Table~\ref{tab-reflection} becomes identical to Table I
of Ref.~\onlinecite{chiu13}.
These ``$\mathbb{Z}_2$'' phases are similar to three-dimensional {\em weak}
$\mathbb{Z}_2$ topological insulators in class AII with (say) two
Dirac cones on a two-dimensional surface, in that the two surface Dirac
cones can be gapped out by a perturbation
(of CDW type) that couples them.
From this analogy we expect that the ``$\mathbb{Z}_2$'' states should be
stable against disorder, if disorder average of any mass term is
assumed to be spatially uniform.
In Appendix \ref{App-reflection},
we discuss the deformation of ``$\mathbb{Z}_2$'' to 0
in more detail
in terms of Clifford algebras.
We will also show in Appendix \ref{App-stability} that 
the deformation of non-trivial states with CDW-type perturbation 
takes place only for these second descendants ``$\mathbb{Z}_2$''
of the $R^-$ and $R^{--}$ cases
and no other such deformations are possible
in Table \ref{tab-reflection}.

\section{multiple additional symmetries\label{sec:multiple}}
We generalize the analysis of Sec.~\ref{sec:additional}
to systems with multiple additional symmetries $\{M_i\}$.
Here, we only consider the situation where
additional symmetries anti-commute with each other, 
\begin{equation}
\{M_i,M_j \}=2\delta_{i,j},
\end{equation}
as well as with zero-dimensional Hamiltonian (Dirac mass) $H$,
$\{M_i,H\}=0$.
The dimensional shift discussed in Sec.~\ref{sec:dimensional shift} is
a special case of choosing $M_i=\gamma_i$.
Other interesting applications can be found in
systems with several independent reflection symmetries ($R_i$) along
different directions. 
Since independent $R_i$'s should commute with each other,
the additional symmetries $M_i=i\gamma_i R_i$ constructed from $R_i$
anti-commute with each other,
and the results of this section are applicable.

\subsection{Complex classes}
The consequences of imposing multiple additional symmetries
on systems in complex classes are as follows.

Class A: The index $q$ of the classifying space $C_q$ is shifted by +1
each time an additional symmetry is imposed, so that the relevant classifying
space becomes $C_N$ when the number of $M_i$'s imposed is $N$.

Class AIII: Suppose that the $M_i$'s have the following algebraic
relations with the chiral symmetry operator $\Gamma$:
\begin{equation}
M_i\Gamma=\left\{\begin{array}{ll}
-\Gamma M_i, & 1\le i\le m,\\
+\Gamma M_i, & m+1\le i\le m+n.
\end{array}\right.
\end{equation}
We then define new generators as $e_0=H$, $e_1=\Gamma$, 
$e_i^-=M_i$ ($i=1,\ldots,m$), and $e_i^+=\Gamma M_{i+m}$ ($i=1,\ldots,n$),
such that
$\{e_0, e_1, e_1^-, \ldots, e_m^-\}$ and 
$\{e_1^+,\ldots, e_n^+\}$ form two Clifford algebras
which commute with each other.
Thus we have an extension problem
$Cl_{m+1}\tensor Cl_{n} \to Cl_{m+2}\tensor Cl_{n}$,
for which the classifying space is $C_{m+1}$.

\subsection{Real classes}
We separately discuss the cases where either of TRS or PHS is present and
the cases where both are present.

(i) $T$ only (AI and AII) or $C$ only (C and D):
Let us write the symmetry operators $M_i$
as $M_i^{\eta_T}$ or $M_i^{\eta_C}$.
with a superscript indicating its signature $\eta_T$ or $\eta_C$
defined in Eq.\ (\ref{eq:TMandCM}).
We denote the numbers of
$M_i^+$'s and $M_i^-$'s by $n^+$ and $n^-$, respectively.
Now we can construct a new Clifford algebra with the generators
\begin{equation}
\{e_0,e_1,e_2,\tilde e^+_1,\ldots,\tilde e^+_{n^+},
 \tilde e^-_1,\ldots,\tilde e^-_{n^-}\},
\end{equation}
where $e_0$, $e_1$, and $e_2$ are defined in Eqs.~(\ref{generators-Tonly})
and (\ref{generators-Conly}), and $\tilde e^\pm_i$ are defined by
\begin{equation}
\tilde e^+_i=JM^+_i, \qquad
\tilde e^-_i=M^-_i,
\end{equation}
as in Table~\ref{tab-M-real}.
We thus find that, upon imposing additional symmetries $M_i^\pm$,
the relevant classifying space is changed from $R_q$ to $R_{\tilde q}$ with
\begin{equation}
\tilde q=
\begin{cases}
q+ n^+ - n^- & (T \t{ only: AI and AII}), \\
q+ n^- - n^+ & (C \t{ only: C and D}).
\end{cases}
\end{equation}

(iii) Both $T$ and $C$ (BDI, DIII, CII, and CI):
We use the notation $M_i^{\eta_T\eta_C}$ for the additional symmetries
with the signatures $\eta_T,\eta_C=+$ or $-$ specifying commutation
or anti-commutation relations with $T$ and $C$.
We denote the numbers of $M_i^{\eta_T \eta_C}$'s by $n^{\eta_T \eta_C}$. 
We then define generators as
\begin{subequations}
\begin{eqnarray}
\tilde e^{+-}_i\!\!&=&\!\!M_i^{+-} \quad (i=1,\ldots,n^{+-}),\\
\tilde e^{-+}_i\!\!&=&\!\!JM_i^{-+} \quad (i=1,\ldots,n^{-+}),\\
\tilde e^{++}_i\!\!&=&\!\!TCM_i^{++} \quad (i=1,\ldots,n^{++}),\\
\tilde e^{--}_i\!\!&=&\!\!TCJM_i^{--} \quad (i=1,\ldots,n^{--}),
\end{eqnarray}
\end{subequations}
with which we have two decoupled Clifford algebras,
\begin{subequations}
\begin{equation}
\{e_0,e_1,e_2,e_3,\tilde e^{+-}_1,\ldots,\tilde e^{+-}_{n^{+-}},
 \tilde e^{-+}_1,\ldots,\tilde e^{-+}_{n^{-+}}\}
\end{equation}
and
\begin{equation}
\{\tilde e^{++}_1,\ldots,\tilde e^{++}_{n^{++}},
 \tilde e^{--}_1,\ldots,\tilde e^{--}_{n^{--}}\}.
\end{equation}
\end{subequations}
For each symmetry class, the generators $\{e_0,e_1,e_2,e_3\}$ 
are given in Eq.\ (\ref{generators-TandC}),
and the original extension problem $Cl_{p,q}\to Cl_{p,q+1}$
(in zero dimension)
is listed in Table \ref{tab-real-AZ}.
With the additional generators, a new extension problem is dictated as
\begin{equation}
Cl_{p+n^{-+},q+n^{+-}} \tensor Cl_{m_1,m_2} \to
 Cl_{p+n^{-+},q+n^{+-}+1} \tensor Cl_{m_1,m_2}
\label{extension-multiple}
\end{equation}
with $(m_1,m_2)=(n^{++},n^{--})$
for DIII and CI, and $(n^{--},n^{++})$ for BDI and CII.
We can show by using Eq.\ (\ref{p+1,q+1}) and dropping $\mathbb{R}(2)$
that the extension (\ref{extension-multiple}) is equivalent to
\begin{align}
Cl_{0, \tilde q} \tensor Cl_{0,m} \to Cl_{0, \tilde q+1} \tensor Cl_{0,m}
\label{extension-multiple-2}
\end{align}
with
\begin{subequations}
\label{eq-tilde-q-m}
\begin{align}
\tilde q&=q+n^{+-}-p-n^{-+},&\\
 m&=
\begin{cases}
n^{--}-n^{++} & (\t{DIII and CI}), \\
n^{++}-n^{--} & (\t{BDI and CII}).
\end{cases}
\end{align}
\end{subequations}
We see that $n^{+-}$ and $n^{-+}$ cause 
a shift in $\tilde q$ of a relevant classifying space,
which can be understood from successive applications of 
the procedure described in Sec.\ \ref{sec:additional}.
On the other hand, $n^{--}$ and $n^{++}$
may effectively alter the base field of the algebra from real to 
complex or quaternion, according to the value of $m$;
see Table~\ref{tab-clifford}(b).
This change cannot be fully expected from
successive applications of 
the procedure in Sec.\ \ref{sec:additional}.
Changing the base field into the complex numbers brings
real symmetry classes to complex ones ($m=3,7$ mod 8),
while changing the base field into the quaternion shifts
the relevant classifying space by $4$ in the periodic table,
Table~\ref{tab-bott}(b) ($m=4,5,6$ mod 8).

\begin{table}
\begin{center}
\caption{
Classification with multiple additional symmetries 
for classes with both TRS and PHS (BDI, DIII, CII, and CI). 
When the numbers of additional symmetries $M^{\eta_T\eta_C}$ are
given by $n^{\eta_T\eta_C}$,
the relevant Clifford algebras and classifying spaces
can be read off from the indices $m$ and $\tilde q$
defined in Eqs.~(\ref{eq-tilde-q-m})
The second column shows Clifford algebras to be extended
as $\tilde q \to \tilde q+1$,
Eq.~(\ref{extension-multiple-2}).
The last column shows the classifying space,
whose zero-th homotopy group gives topological
classification.
}
\begin{tabular}{cccc}
\hline \hline 
$m$ (mod 8) & Clifford algebra & classifying space \\
\hline
$0 $ & $ Cl_{0,\tilde q} $ & $ R_{\tilde q} $ \\
$1 $ & $ Cl_{0,\tilde q} \oplus Cl_{0,\tilde q} $
 & $ R_{\tilde q} \times R_{\tilde q} $ \\
$2 $ & $ Cl_{0,\tilde q} $ & $ R_{\tilde q} $ \\
$3 $ & $ Cl_{\tilde q} $ & $ C_{\tilde q} $ \\
$4 $ & $ Cl_{0,\tilde q+4} $ & $ R_{\tilde q+4} $ \\
$5 $ & $ Cl_{0,\tilde q+4} \oplus Cl_{0,\tilde q+4} $
 & $ R_{\tilde q+4} \times R_{\tilde q+4} $ \\
$6 $ & $ Cl_{0,\tilde q+4} $ & $ R_{\tilde q+4} $ \\
$7 $ & $ Cl_{\tilde q} $ & $ C_{\tilde q} $ \\
\hline \hline
\end{tabular}
\label{tab-multiple}
\end{center}
\end{table}

The new classification arising from the extension problem
(\ref{extension-multiple-2}) is summarized in Table~\ref{tab-multiple}
(some mathematical details needed in the derivation can be found
in Appendix~\ref{appendix-mult}).
With Table \ref{tab-multiple}, we can find topological classification
in the presence of additional symmetries as follows.
First, without additional symmetries, we identify the relevant
Clifford algebra $Cl_{p,q}$
for a given symmetry class in Table \ref{tab-real-AZ}.
Next from the (anti-)commutation relations of additional symmetries $M_i$'s
with $T$ and $C$, we find the numbers $n^{\eta_T \eta_C}$,
which determine $\tilde q$ and $m$ in Eqs.\ (\ref{eq-tilde-q-m}).
Table~\ref{tab-multiple} then tells us associated extension problem
of the relevant Clifford algebra.
Once the relevant classifying space is found,
the topological classification of zero-dimensional systems is obtained
from the zeroth homotopy group, which is listed in Table \ref{tab-bott}.
As for $d$-dimensional systems,
we only have to replace $\tilde q$ with $\tilde q - d$,
as we have discussed in Sec.~\ref{sec:dimensional shift}
(each gamma matrix in the kinetic term of Dirac Hamiltonian gives $M^{-+}$).
For a classifying space of a direct product such as $R_q \times R_q$,
the $d$-dimensional topological invariant is also a direct product
as $\pi_0(R_{q-d}) \times \pi_0(R_{q-d})$.

Finally, let us point out an interesting consequence from
Table \ref{tab-multiple}.
As we decrease $m$ from 0 to $-2$ by imposing some additional symmetries,
we find that the classification changes into a complex class at $m=-1$,
and at $m=-2$ it comes back to a real symmetry class with a shift
of 4 in the Bott periodicity. 
We will discuss an example of this type in Sec.~\ref{sec:examples},
where two types of reflection symmetries are imposed
to cause such changes in a symmetry class.

\section{examples\label{sec:examples}}
We discuss several examples of insulators or superconductors
which exhibit a change in topological properties
due to addition of reflection or mirror symmetries.
We will use ``$\, i\,$'' in real algebras instead of $J$ in this section.

\subsection{Mirror Chern number}
In relation to mirror Chern numbers\cite{teo-fu-kane08} which
characterize a recently discovered topological crystalline
insulator SnTe,\cite{hsieh12,xu2012observation,tanaka2012experimental,dziawa2012topological}
let us consider Dirac Hamiltonian defined by
\begin{align}
H=m \sigma_z + v (k_x s_y-k_y s_x)\sigma_x+v_z k_z \sigma_y,
\label{eq-snte}
\end{align}
where $\sigma_i$ and $s_i$ are Pauli matrices for orbital and spin
degrees of freedom.
This Hamiltonian\cite{hsieh12} describes low-energy excitations
near an $L$ point of SnTe
and belongs to class AII, since it is invariant under
time-reversal operation, $T=is_y\mathcal{K}$.
We can define a topological index which is determined by the sign of $m$.
If we double the degrees of freedom to account for another $L$ point as
$H_2=H\otimes\tau_0$, where $\tau_0$ is a unit $2\times2$ matrix in
the valley sector ($L_1$ and $L_2$), then the Hamiltonian can have
another $T$-invariant mass term,
$m's_z\sigma_x\tau_y$,
where $\tau_y$ is a Pauli matrix and couples the two valleys.
With the additional mass $m'$,
insulators with different signs of $m$ are
no longer topologically distinguished, because rotation in
the $m$-$m'$ plane can adiabatically connect the two.
Thus $H$ is characterized by a $\mathbb{Z}_2$ topological index.

Let us now take into consideration reflection symmetry
in the $x$ direction ($k_x\to-k_x$).
The Hamiltonian $H$ is indeed invariant under the reflection transformation,
which can be written as $R^{-1}H(-k_x,k_y,k_z)R=H(k_x,k_y,k_z)$ with $R=s_x$.
Following the analysis in Sec.~\ref{sec:additional}B,
we define $M=is_y\sigma_xs_x=\sigma_xs_z$,
which anticommutes with $T$, i.e., $\eta_T=-1$.
We then find from Table~\ref{tab-reflection} that class AII with
the reflection $R^-$ effectively behaves like class DIII,
which is characterized by an integer topological number $\mathbb{Z}$ in $d=3$.
As we discuss below,
the integer topological number corresponds to the mirror Chern
number.\cite{teo-fu-kane08,hsieh12}
Incidentally, the mirror symmetry does not allow the doubled Hamiltonian
$H_2=H\otimes\tau_0$ to have the additional mass term $m's_z\sigma_x\tau_y$,
so that the topological numbers from the two valley sectors can add up.

The mirror Chern number is a Chern number defined from Bloch
states on the mirror plane ($k_x=0$) for each eigenspace of
the reflection $R$ (i.e., $s_x=\pm 1$).
In each subspace the Hamiltonian (\ref{eq-snte}) reduces to
\begin{align}
H^\pm=m \sigma_z \mp v k_y \sigma_x+v_z k_z \sigma_y,
\label{eq-snte-mirror}
\end{align}
which falls into class D (with $C=\sigma_x\mathcal{K}$)
and apparently possesses
a Chern number $\mathbb{Z}$.
Now let us relate this mirror Chern number
to our classification.
The original extension problem for class AII with mirror symmetry
in $d=3$ is
$Cl_{2,4} \to Cl_{3,4}$,
which we find from Eq.\ (\ref{eq-app-Cl-R}) is equivalent to 
$Cl_{2,2} \tensor Cl_{0,2} \to Cl_{2,3}\tensor Cl_{0,2}$.
The latter can be regarded as an extension problem for class D in $d=2$,
if we drop the trivial part $Cl_{0,2} \simeq \mathbb{R}(2)$.
The original generators of the Clifford algebra for class AII
with a mirror forming $Cl_{3,4}$ are
$\{ e_1, e_2, \gamma_x,\gamma_y,\gamma_z, M, e_0  \}$,
where $(e_1,e_2,e_0)=(is_y\mathcal{K},s_y\mathcal{K},i\sigma_z)$ are defined in
Eq.\ (\ref{generators-Tonly}),
and $(\gamma_x,\gamma_y,\gamma_z)=(s_y \sigma_x, -s_x\sigma_x,\sigma_y)$ 
come from kinetic terms in $H$.
Using these generators, we can construct generators for the new algebra
$Cl_{2,3}\tensor Cl_{0,2}$
explicitly as
\begin{align}
&\{ -e_2 \gamma_x, e_1 \gamma_x, \gamma_x \gamma_y M, -e_1 e_2\gamma_z,
 e_1 e_2 e_0  \} 
\n &
\tensor \{ e_1e_2\gamma_x M, e_1e_2\gamma_x \gamma_y \} \n
&= \{ \sigma_x \mathcal{K}, -i \sigma_x \mathcal{K},
      i \sigma_x, i \sigma_y, \sigma_z \} \tensor
\{ s_x, s_z\}.
\label{generators for D}
\end{align}
The latter half of the right-hand side spans the spin degrees of freedom $s_i$,
while the former half corresponds to a Clifford algebra
$Cl_{2,3}$ for Dirac Hamiltonians of class D in $d=2$.
We can read from Eqs.\ (\ref{generators-Conly}) and (\ref{generators for D})
and Table \ref{tab-M-real}(a) that
the particle-hole transformation is $C=\sigma_x \mathcal{K}$,
the gamma matrices in the kinetic terms are $(\sigma_x,\sigma_y)$,
and the mass term $\propto\sigma_z$.
Indeed this construction reproduces the Hamiltonian
in Eq.\ (\ref{eq-snte-mirror}), for which a mirror Chern number is defined.
Thus the topological number $\mathbb{Z}$ for class AII with a mirror in $d=3$ 
is equivalent to a topological number $\mathbb{Z}$ for class D in $d=2$,
i.e., a mirror Chern number.

\subsection{Topological mirror superconductor ($d=1$)}
Next we consider a one-dimensional model of a time-reversal
invariant topological mirror superconductor
discussed in Ref.~\onlinecite{zhang-kane-tmsc},
in which an integer number of Majorana Kramers' pairs live
at an end of a mirror invariant wire.
We consider Hamiltonian of a Rashba quantum wire
with proximity coupling to a nodeless $s_\pm$-wave superconductor,
\begin{eqnarray}
H\!\!&=&\!\!
(-2t \cos k_x+2\lambda \sin k_x \sigma_z -\mu)\tau_z
\n
& &{}\!\! 
+2\Delta_1\tau_x\cos k_x,
\end{eqnarray}
where $t$ is hopping, $\lambda$ Rashba coupling,
$\sigma_i$ and $\tau_i$ are Pauli matrices in the spin and particle-hole
spaces, and the condition $|\mu|<2\lambda$
is assumed.\cite{zhang-kane-tmsc}
This Hamiltonian is in class DIII with $T=i\sigma_y \mathcal{K}$,
$C=\sigma_y\tau_y \mathcal{K}$,
characterized by a $\mathbb{Z}_2$ topological number
and the existence of a Kramers' pair of Majorana zeromodes at the edge.
If we consider a doubled system $H_2=H \tensor \rho_0$ where $\rho_0$ is a 
$2 \times 2$ unit matrix,
we have another mass term $\sigma_x \tau_z \rho_y$ that can deform a 
$\mathbb{Z}_2$ nontrivial state into a trivial state,
indicating that two pairs of Majorana zeromodes are unstable.

Now let us impose a ``mirror line'' symmetry $R=\sigma_z$, which
commutes with $H$.
As we discussed in Sec.~\ref{sec:additional}C,
combining $R$ with the chiral symmetry $TC$, 
we can introduce an additional symmetry $M=\sigma_z\tau_y$,
which anti-commutes with $H$.
Since the commutation relations of $M$ with symmetry operators $T$ and $C$
are characterized by $(\eta_T,\eta_C)=(+1,+1)$,
we have a change in the symmetry class from DIII to AIII
[Table~\ref{tab-M-real}(b)],
so that we can define a topological number $\mathbb{Z}$
in each eigenspace of $R$,
which is a mirror winding number.\cite{zhang-kane-tmsc}
The change in the classification from $\mathbb{Z}_2$ to $\mathbb{Z}$
is reflected in the disappearance of the additional mass term
($\sigma_x \tau_z \rho_y$) in
the doubled system upon imposing $R$,
which guarantees the stability of Kramers' pairs
of Majorana zeromodes.\cite{zhang-kane-tmsc}

\subsection{CI $\to$ AIII $\to$ DIII $(d=2)$}
In this section we construct a toy model in $d=2$
whose symmetry class changes (a) from CI to AIII when
we impose an additional symmetry $M_1$,
and (b) from CI to DIII when we impose two additional symmetries
$M_1$ and $M_2$.
We assume that both symmetry operators $M_1$ and $M_2$ commute
with $T$ and $C$, i.e., $(\eta_T,\eta_C)=(+,+)$.
These symmetries can be thought of as coming from
reflection symmetries in
the $x$ and $y$ directions, for example.
In the context of Sec.~\ref{sec:multiple},
our toy model is an example where
a real symmetry class turns into (a) a complex class
($m=-1$ and $\tilde q=7$ in Table~\ref{tab-multiple})
and (b) a real class shifted by 4 in the Bott period 
($m=-2$ and $\tilde q=7$ in Table~\ref{tab-multiple}).
The $d=2$ topological number of our model changes
as (a) $0\to 0$ and (b) $0\to\mathbb{Z}_2$.

Let us consider a two-dimensional 8 by 8 Dirac Hamiltonian,
\begin{align}
H=(k_x \sigma_x +k_y \sigma_z)\tau_x + \sum_i m_i g_i.
\end{align}
The Dirac mass operators $g_i$ are 8 by 8 matrices made from 3 sets of
Pauli matrices $\sigma$, $\tau$, and $\rho$.
We require the system to be in class CI with the symmetry operators,
\begin{align}
T=\tr yy \mathcal{K}, \qquad
C=\tr xy \mathcal{K}, \qquad
\Gamma=\tau_z,
\label{eq:TCGamma}
\end{align}
which satisfy $T^2=+1$, $C^2=-1$, and $[T,C]=0$.
We find the following set
of possible $g_i$'s which anticommute with the kinetic terms
and which are compatible with the symmetries (\ref{eq:TCGamma}):
\begin{equation}
\tau_y\tensor\{\rho_x,\rho_y,\rho_z\},
\;
\sigma_y\tau_x.
\label{eq:mass terms}
\end{equation}
The two additional symmetries which we are going to impose are
\begin{align}
M_1^{++}=\tr zz, \qquad
M_2^{++}=\tr zy,
\end{align}
both of which anticommute with the kinetic terms and commute with $T$ and $C$.
As discussed in Sec.~\ref{sec:additional} and \ref{sec:multiple},
we can define new operators
$\widetilde M_i=TCM_i^{++}$ which commute with $H$, $T$, and $C$:
$\widetilde M_1=i\rho_z$, $\widetilde M_2=i\rho_y$.

First we impose only the $M_1$ symmetry.
The allowed $g_i$'s which anticommute with $M_1$ are reduced to
\begin{equation}
\tau_y\rho_z,
\;
\sigma_y\tau_x.
\label{g_is with M_1}
\end{equation}
Since $\widetilde M_1=i\rho_z$ commute with $H$, we can concentrate on
the subspace of $\rho_z=+1$.
We then find that the relevant symmetry class is AIII
with the chiral symmetry $\Gamma=\tau_z$.
This is consistent with Table~\ref{tab-M-real}(b), where
the classifying space for class CI
with $(\eta_T,\eta_C)=(+,+)$ and $\widetilde M^2=-1$
is shown to be $C_1$, i.e., class AIII [Table~\ref{tab-bott}(a)].
We observe that,
when $H$ has a mass term taken from Eq.\ (\ref{g_is with M_1}),
we can always add a second mass term which anticommutes with
the first mass term.
This indicates that the system is topologically trivial,
in agreement with topological triviality of class AIII in $d=2$.

When both symmetries $M_1$ and $M_2$ are imposed,
among those in Eq.\ (\ref{eq:mass terms}),
the only allowed Dirac mass operator
which anticommute with both $M_1$ and $M_2$ is
$\sigma_y\tau_x$.
Then the Dirac Hamiltonian reads
\begin{equation}
H=(k_1 \sigma_x +k_2 \sigma_z)\tau_x+m_1 \sigma_y \tau_x,
\label{H DIII}
\end{equation}
which does not contain Pauli matrices $\rho$'s,
since $H$ has to commute with both $\widetilde M_1=i\rho_z$
and $\widetilde M_2=i\rho_y$.
The Hamiltonian (\ref{H DIII}) belongs to class DIII
with new symmetry generators
$T=i\tau_y \mathcal{K}$
and $C=\tau_x \mathcal{K}$.
It has a unique mass term $m_1 \sigma_y \tau_x$
and is characterized by a $\mathbb{Z}_2$ topological index.

\subsection{BDI $\to$ AIII $\to$ CII $(d=3)$}
Next we construct a toy model in $d=3$ whose symmetry class
changes (a) from BDI to AIII when we impose an additional symmetry $M_1$
and (b) from BDI to CII when we impose two additional symmetries $M_1$
and $M_2$.
Accordingly, the $d=3$ topological number changes as
(a) $0 \to \mathbb{Z}$ and (b) $0 \to \mathbb{Z}_2$.

We consider a Dirac Hamiltonian of the form
\begin{align}
H=(k_x \sigma_x s_z + k_y s_x + k_z \sigma_z s_z)\tau_x + \sum_i m_i g_i,
\label{H-BDI}
\end{align}
where the Dirac mass operators $g_i$'s are 16 by 16 matrices
written as products of Pauli matrices 
$\sigma$, $s$, $\tau$, and $\rho$.
We choose basic symmetry operators as
\begin{equation}
T=s_y\tr 0y \mathcal{K}, \qquad
C=s_y\tr zy \mathcal{K}, \qquad
\Gamma=\tau_z,
\label{eq:TCGamma2}
\end{equation}
which makes the model to be in class BDI ($T^2=+1$, $C^2=+1$).
The Dirac mass operators $g_i$'s, which anticommute with the
kinetic terms in Eq.\ (\ref{H-BDI}) and are compatible with
the symmetries (\ref{eq:TCGamma2}), are taken from the following set:
\begin{equation}
\sigma_y s_z \tau_x, \;
\{s_y\tau_x, \tau_y, \sigma_y s_x \tau_y\}\tensor\{\rho_x, \rho_y, \rho_z\}.
\end{equation}
The additional symmetries to be imposed are again given by
\begin{align}
M_1^{--}=\tr zz, \qquad
M_2^{--}=\tr zy,
\end{align}
which anticommute with $T$ and $C$,
i.e., $(\eta_T,\eta_C)=(-,-)$.
We then define
$\widetilde M_i=iTCM_i^{--}$, which commute with $H$, $T$, and $C$:
$\widetilde M_1=i\rho_z$, $\widetilde M_2=i\rho_y$.
We may think of $M_1^{--}$ and $M_2^{--}$ as coming from
reflection symmetries along the $x$ and $y$ directions,
respectively.

Let us impose only the $M_1^{--}$ symmetry.
The allowed $g_i$'s anticommuting with $M_1^{--}$ are reduced to
\begin{equation}
\sigma_y s_z \tau_x, \;
\{s_y\tau_x, \tau_y, \sigma_y s_x \tau_y\}\tensor\rho_z.
\label{eq:mass terms-2}
\end{equation}
Since $[\rho_z,H]=0$, we can concentrate on the eigenspace $\rho_z=+1$ (or $-1$)
and find that the system in this subspace is in class AIII with
the chiral symmetry $\Gamma=\tau_z$.
This is in agreement with Table~\ref{tab-M-real}(b), where
the classifying space for class BDI with $(\eta_T,\eta_C)=(-,-)$ is
shown to be $C_1$,
as well as with Table~\ref{tab-multiple} ($m=-1$ and $\tilde q=1$).
We observe that, among the mass operators in Eq.\ (\ref{eq:mass terms-2}),
$\sigma_ys_x\tau_y\rho_z$ is special in that it commutes with the other
three operators, while these three anticommute among themselves.
This means that Hamiltonians with different signs of the mass term
$\sigma_ys_x\tau_y\rho_z$ are topologically distinct,
which is consistent with the fact that systems in class AIII
are characterized by an integer topological index.

When both symmetries $M_1$ and $M_2$ are imposed,
we have only a single allowed mass operator, $\sigma_ys_z\tau_x$.
In this case the Hamiltonian has the form
\begin{equation}
H=(k_x \sigma_x s_z + k_y s_x + k_z \sigma_z s_z)\tau_x+m\sigma_ys_z\tau_x,
\end{equation}
which turns out to be in class CII with 
$T=s_y \mathcal{K}$ and $C=s_y \tau_z \mathcal{K}$,
and is characterized by a $\mathbb{Z}_2$ topological index.
The change in the symmetry class from BDI to CII is indeed expected from
Table \ref{tab-multiple} with $m=-2$ and $\tilde q=1$ indicating
the classifying space to be $R_5$, which corresponds to class CII
(Table~\ref{tab-bott}).

\section{Summary}
We have studied changes in classification of topological insulators
and superconductors due to additional symmetries,
by considering extension problems of Clifford algebras generated
from operators representing symmetry constraints.
Our theory provides a simple and clear derivation of topological
classification which agrees with the periodic table
obtained by Chiu \textit{et al.},\cite{chiu13}
who studied topological invariants for topological
insulators and superconductors with a reflection symmetry.
We have also discussed several examples including a topological
crystalline insulator characterized by mirror Chern numbers
and mirror topological superconductors.

\section{acknowledgment}
This work was supported by Grant-in-Aid from the Japan Society for
Promotion of Science (Grant Numbers 24840047 and 24540338)
and by the RIKEN iTHES Project.

\appendix
\section{Properties of the Clifford algebras\label{appendix-Cl}}

\begin{table}
\begin{center}
\caption{
Clifford algebras and classifying spaces for (a) complex and (b) real classes
(after Table 2 of Ref.~\onlinecite{kitaev09}).
The last columns denote the zero-th homotopy group of each classifying space.
}

(a) complex classes
\\
\begin{tabular}[t]{cccc}
\hline \hline
$q$ & $Cl_q$ & $C_q$ & $\pi_0(C_q)$ \\
\hline 
0 & $\mathbb{C}$ & $(U(n+m)/U(n)\times U(m))\times \mathbb{Z}$
  & $\mathbb{Z}$ \\  
1 & $\mathbb{C} \oplus \mathbb{C}$ & $U(n)$ & 0 \\  
\hline \hline
\end{tabular}
\\ \vspace{1em}

(b) real classes
\\
\begin{tabular}[t]{cccc}
\hline \hline
$q$ & $Cl_{0,q}$ & $R_q$ & $\pi_0(R_q)$ \\
\hline 
0 & $\mathbb{R}$ & $(O(n+m)/O(n)\times O(m))\times \mathbb{Z}$
  & $\mathbb{Z}$ \\  
1 & $\mathbb{R} \oplus \mathbb{R}$ & $O(n)$ & $\mathbb{Z}_2$ \\  
2 & $\mathbb{R}(2)$ & $O(2n)/U(n)$ & $\mathbb{Z}_2$ \\  
3 & $\mathbb{C}(2)$ & $U(2n)/Sp(n)$ & $0$ \\  
4 & $\mathbb{H}(2)$ & $(Sp(n+m)/Sp(n)\times Sp(m))\times \mathbb{Z}$
  & $\mathbb{Z}$ \\  
5 & $\mathbb{H}(2) \oplus \mathbb{H}(2)$ & $Sp(n)$ & $0$ \\  
6 & $\mathbb{H}(4)$ & $Sp(n)/U(n)$ & $0$ \\  
7 & $\mathbb{C}(8)$ & $U(n)/O(n)$ & $0$ \\  
\hline \hline
\end{tabular}
\label{tab-clifford}
\end{center}
\end{table}

We summarize some useful formulas of Clifford algebras
used in this paper.
We begin with complex Clifford algebras:
\begin{subequations}
\begin{align}
Cl_{1} &\simeq \mathbb{C} \oplus \mathbb{C} , \\
Cl_{2} &\simeq \mathbb{C}(2), \\
Cl_{n+2} &\simeq Cl_{n} \tensor \mathbb{C}(2).
\end{align}
\label{eq-app-Cl-C}%
\end{subequations}
The classifying space for the extension problem $Cl_n \to Cl_{n+1}$
is denoted by $C_n$.
Since a fixed representation for 2 by 2 complex matrices $\mathbb{C}(2)$
does not affect the extension problem,
Eq.\ (\ref{eq-app-Cl-C}c) leads to a periodic structure
of the classifying space $C_{n+2}\simeq C_n$.

The real Clifford algebras have the following properties:\cite{karoubi}
\begin{subequations}
\begin{align}
Cl_{0,1} &\simeq \mathbb{R} \oplus \mathbb{R} , \\
Cl_{0,2} &\simeq \mathbb{R}(2), \\
Cl_{1,0} &\simeq \mathbb{C}, \\
Cl_{2,0} &\simeq \mathbb{H}, \\
Cl_{p+1,q+1} &\simeq Cl_{p,q} \tensor \mathbb{R}(2), \label{p+1,q+1}\\
Cl_{p,q} \tensor Cl_{0,2} &\simeq Cl_{q,p+2}, \\
Cl_{p,q} \tensor Cl_{2,0} &\simeq Cl_{q+2,p}, \\
Cl_{p,q} \tensor Cl_{0,4} &\simeq Cl_{p,q+4},
\end{align}
\label{eq-app-Cl-R}%
\end{subequations}
where $\mathbb{H}$ denotes the set of quaternions.
The classifying space
for the extension problem $Cl_{p,q} \to Cl_{p,q+1}$ is given by $R_{q-p}$.
The extension problems
$Cl_{p,q} \tensor \mathbb{R}(2)\to Cl_{p,q+1}\tensor \mathbb{R}(2)$
and $Cl_{p,q}\to Cl_{p,q+1}$ are equivalent,
since the sector of 2 by 2 real matrices $\mathbb{R}(2)$ with
fixed representation does not change the degrees of freedom
of the extension.
Furthermore, we have 
\begin{equation}
Cl_{p+8,q} \simeq Cl_{p,q+8} \simeq Cl_{p,q} \tensor \mathbb{R}(16).
\end{equation}
Since $\mathbb{R}(16)$ does not affect extension problems,
the classifying space $R_{q}$ has the periodic structure $R_{q+8}\simeq R_q$.

Clifford algebras and classifying spaces for complex and real classes
are summarized in Table~\ref{tab-clifford}.

\section{Deformation of second descendant $\mathbb{Z}_2$ into 0 under $R^-$
or $R^{--}$ symmetry \label{App-reflection}}
In this appendix we show, first with an explicit example and second
by applying Clifford algebras for general cases, 
that insulating states characterized with a non-trivial
second descendant $\mathbb{Z}_2$ index [$=\!\pi_0(R_2)$]
in the presence of a reflection symmetry $R^{--}$ (or $R^-$),
can be deformed into a topologically trivial state with a mass term that
mixes Dirac cones at different momenta,
when the translation symmetry is not assumed.

Let us begin with an example of a two-dimensional 8 by 8 Hamiltonian
written as
\begin{align}
H=\sigma_x \tau_z k_x + \sigma_y \tau_z k_y + \sum_i m_i g_i,
\end{align}
which we assume to be in class CII with the symmetry operators
\begin{align}
T&=i\sigma_y \rho_z \mathcal{K}, &
C&=i\sigma_y\tau_x\rho_z\mathcal{K},&
\Gamma&=\tau_x.
\label{TCGamma}
\end{align}
The Dirac mass terms $g_i$'s, which anti-commute with kinetic terms and 
are compatible with the generic symmetries in Eq.\ (\ref{TCGamma}),
are given by
\begin{align}
g_1=\sigma_z\tau_z\rho_x,& ~~~  g_2=\tau_y\rho_x.
\label{app-ref-z2-mass}
\end{align}
The presence of these two mutually anticommuting mass terms implies
that the system is topologically trivial ``0''.
Now we impose a reflection symmetry along the $x$ direction,
\begin{align}
R^{--}=\sigma_y\rho_z,
\end{align}
which anti-commutes with $T$ and $C$ defined in Eq.\ (\ref{TCGamma}).
Then the possible Dirac mass term is $g_1$ only,
and the system is classified by a $\mathbb{Z}_2$ index.

So far we implicitly assumed translation symmetry.
However, if we admit terms breaking the translation symmetry,
we can find an extra mass term that anti-commutes with $g_1$
in Eq.\ (\ref{app-ref-z2-mass})
and turns a $\mathbb{Z}_2$ topological state into a trivial
state, as we explain below.
First, we add to the Hamiltonian a term
$\delta\Delta=\delta\sigma_x\tau_z\rho_x$.
Since it commutes with the kinetic term in the $x$ direction,
it causes splitting of the Dirac points
as $k_x \to k_x \pm \delta$.
We then add an extra mass term $m_2(x) \tau_y\rho_x$,
which is reflection symmetric if $m_2(x)$ is an odd function of $x$.
Now the Hamiltonian reads
\begin{align}
H=\sigma_x \tau_z (k_x +\delta \rho_x) + \sigma_y \tau_z k_y +
m_1 \sigma_z\tau_z\rho_x + m_2(x) \tau_y\rho_x.
\end{align}
Let the Fourier component of $m_2(x)$ at $k_x=2\delta$ be $m_2^\delta$.
Then rotations in the masses $(m_1,m_2^\delta)$ will connect any insulating
state to trivial insulators.
In particular, gapless edge states, formed along the boundary parallel
to the $x$ axis, will be gapped out by the $m_2$ term with a finite
$m_2^\delta$.
Hence we have only a single (trivial) insulating phase.

The disappearance of the second descendant $\mathbb{Z}_2$ topological states
can be also understood in terms of Clifford algebras.
The deformation to a trivial state is possible when
we have the two kinds of operators $\Delta$ and $g_2$
with the following properties:
\begin{itemize}
\item
$\Delta$ commutes with $\gamma_x$, anti-commutes with
the other kinetic terms and the mass term ($g_1$), and
is compatible with TRS, PHS, and $R^{--}$.
\item
$g_2$ anti-commutes with $g_1$, $\Delta$, and the kinetic terms,
and is compatible with TRS and PHS, but {\em not} with $R^{--}$.
\end{itemize}
When these two operators are available, we can split Dirac points along the
$k_x$ direction with $\Delta$, and couple
the split Dirac points by adding an extra mass term
$m_2(x)g_2$ with the mass modulation $m_2(x)$ which is odd in $x$.

The existence condition for a $\Delta$-type operator can be formulated
as an extension problem of Clifford algebra.
The Clifford algebra characterizing class CII
in $d=2$ with $R^{--}$ is given by
\begin{align}
\{e_0, e_1, e_2, e_3, J\gamma_x, J\gamma_y, \gamma_x R^{--} \},
\label{generators CII with R--}
\end{align}
where
$e_i$'s are defined in Eq.\ (\ref{generators-TandC}), and
the extension $Cl_{6,0}\to Cl_{6,1}$ yields a classifying space $R_2$.
If we have an operator $\tilde \Delta$ that
squares to $-1$ and anti-commutes with all the
generators in Eq.\ (\ref{generators CII with R--}),
then we can define $\Delta$ as
\begin{align}
\Delta=J\tilde\Delta R^{--}.
\end{align}
Thus the problem which we need to consider is
whether the Clifford algebra $Cl_{5,1}$, generated by
$\{e_0, e_1, e_2, e_3, J\gamma_y, \gamma_x R^{--} \}$,
can be extended by introducing
another generator squaring to $-1$ (i.e., $J\gamma_x$ or $\tilde \Delta$). 
The extension is $Cl_{5,1}\to Cl_{6,1}$ with a classifying space $R_6$,
whose topological index is ``0''.
This ensures simultaneous existence of $J\gamma_x$ and $\tilde \Delta$ that 
anti-commute with each other, hence, the existence of $\Delta$.
This argument can be applied to other spatial dimensions as well.
When the original classification of a given
second descendant ``$\mathbb{Z}_2$'' state is characterized by
$Cl_{p,q} \to Cl_{p,q+1}$ (corresponding classifying space, $R_{q-p}$),
the extension problem regarding $\tilde \Delta$ is
$Cl_{p-1,q+1} \to Cl_{p,q+1}$,
whose classifying space is $R_{p-q}$.
Since we have $q-p=2$ (mod 8) for the second descendant $\mathbb{Z}_2$ states, 
we can always find a $\Delta$ operator because $\pi_0(R_{p-q})=\pi_0(R_6)=0$.

Next, we repeat a similar discussion to show the existence of 
a $g_2$-type operator.
If we have an operator $\tilde g_2$ that squares to $-1$ and
anti-commutes with all the generators in the Clifford algebra
\begin{align}
\{e_0, e_1, e_2, e_3, J\gamma_x, J\gamma_y,
   \gamma_x R^{--}, -J\Delta R^{--} \},
\label{generators CII with R-- and Delta}
\end{align}
we can construct $g_2$ as
\begin{align}
g_2=J \Delta \gamma_x \tilde g_2.
\end{align}
The existence of $\tilde g_2$ is established by considering 
the extension problem $Cl_{p,q+1}\to Cl_{p+1,q+1}$, 
where we try to add $\tilde \Delta$ to (\ref{generators CII with R--}).
The topology of the associated classifying space
$\pi_0(R_{p-q+1})=\pi_0(R_{7})=0$ 
tells that we always have another generator anti-commuting
with $\tilde \Delta$, namely, $\tilde g_2$.
Thus we find that a $g_2$-type operator also exists.
The existence of the two types of operators ($g_2$ and $\Delta$)
reduces the second descendant $\mathbb{Z}_2$ states
to trivial ``0'' states.

A similar argument can be applied to the second descendant $\mathbb{Z}_2$
states under a $R^-$ symmetry, and we conclude that ``$\mathbb{Z}_2$''
in Table~\ref{tab-reflection} are changed into 0,
when we include non-uniform terms like $m_2(x)g_2$.

Finally, we dicuss stability of the surface states of a ``$\mathbb{Z}_2$''
nontrivial insulator with $R^{--}$ or $R^-$ symmetry.
We have shown above that the surface states are
gapless if $m_2^\delta=0$.
Now we ask what happens if we take $m_2(x)$ to be a random odd function of $x$.
We argue below that the surface states remain critical,
as long as random perturbation is spatially uniform on average
(disorder average of $m_2^\delta$ vanishes).
To this end, we consider classification of the $g_2$-type mass term.
Let us examine the extension problem of (\ref{generators CII with R-- and Delta})
with $\tilde g_2$.
The extension is characterized with
$Cl_{p+1,q+1} \to Cl_{p+2,q+1}$,
whose classifying space turns out to be $R_{p-q+2}$.
This means that the $g_2$-type operator is 
classified with $\pi_0(R_0)=\mathbb{Z}$.
For slow modulation of $m_2(x)$ we can imagine that
the surface is divided into domains possessing various values of $\mathbb{Z}$,
with gapless edge states running along domain boundaries and percolating
through the surface. 
Thus we can expect that the gapless surface states
of a ``$\mathbb{Z}_2$'' topological insulator
remain critical, when random perturbations are spatially uniform on average.
This is similar to criticality of surface states of
3D weak topological insulators in the presence of random
potential with zero mean.\cite{Mong,Ringel,FuKane2012,fulga2012statistical}

\section{Stability of topological states with reflection symmetry \label{App-stability}}
In Appendix \ref{App-reflection},
we have shown that the second descendant $\mathbb{Z}_2$ states
with $R^{--}$ or $R^-$ reflection symmetry
(labeled with ``$\mathbb{Z}_2$'' in Table \ref{tab-reflection}) 
can be deformed into trivial states
with a CDW type mass term.
Here we prove that, besides the second descendant ``$\mathbb{Z}_2$''
states, there are no such deformations of topological states to trivial ones
in Table \ref{tab-reflection}.

For simplicity we concentrate on systems with both TRS and PHS,
for which we have the Clifford algebra
\begin{align}
\{e_0,e_1,e_2,e_3,J\gamma_1,J\gamma_2,\ldots,J\gamma_d\},
\end{align}
supplemented by either $\tilde e$ or $\widetilde{M}$ in Table \ref{tab-M-real}
related to the reflection symmetry $M=J \gamma_1 R$.

To deform a topologically non-trivial state into a trivial state
by introducing a CDW-type mass perturbation,
we need have two terms $\Delta$ and $g_2$ satisfying the following
conditions:
\begin{itemize}
\item
$\Delta$ commutes with $\gamma_1$, anti-commutes with
the other $\gamma_i$'s and the mass term ($e_0$),
and is compatible with generic symmetries (TRS, PHS and chiral)
and the reflection symmetry.
\item
$g_2$ anti-commutes with $e_0$, $\Delta$ and all the $\gamma_i$'s,
and is compatible with generic symmetries,
but {\em not} with the reflection symmetry.
\end{itemize}

In general, 
the existence of an operator $e_+$ ($e_-$) which squares to $+1$ ($-1$)
and which can be used as a new generator to extend
a Clifford algebra $Cl_{p,q}$,
is judged by considering
an extension problem of an algebra with one generator fewer.
Namely, we look at an extension problem 
$Cl_{p,q-1} \to Cl_{p,q}$ for $e_+$
($Cl_{p-1,q} \to Cl_{p,q}$ for $e_-$).
If topological classification of the associated classifying space
is ``0'', the existence of $e_{\pm}$ is guaranteed.
On the other hand,
if the classification is $\mathbb{Z}$ or $\mathbb{Z}_2$,
we cannot have $e_{\pm}$.
We note that this last statement is valid under the assumption that
we are dealing with minimal models where Hamiltonian is already
block diagonalized with respect to possible commuting unitary
operators.\cite{note}

\subsection{$R^{--}$ case}
We assume that the original classification for the mass term $e_0$ is
given by the extension $Cl_{p,q} \to Cl_{p,q+1}$,
where $Cl_{p,q+1}$ has generators
\begin{align}
\{e_0,e_1,e_2,e_3,J\gamma_1, \ldots, J\gamma_d, \gamma_1 R^{--} \},
\end{align}
whose classifying space is $R_{q-p}$.
The question as to whether the $\Delta$-type operator can be added
to the algebra as
\begin{align}
\{e_0,e_1,e_2,e_3,J\gamma_1,\ldots,J\gamma_d, \gamma_1 R^{--}, J\Delta R^{--} \},
\end{align}
is answered by examining the extension problem 
$Cl_{p-1,q+1} \to Cl_{p,q+1}$, because $(J\Delta R^{--})^2=-1$.
We find that the classifying space is $R_{p-q}$.
As for a $g_2$-type operator, the extended algebra with $g_2$
is written as
\begin{align}
\{e_0,e_1,e_2,e_3,J\gamma_1,\ldots,J\gamma_d, \gamma_1 R^{--},
 J\Delta R^{--}, J\Delta \gamma_1 g_2 \},
\end{align}
for which the extension is $Cl_{p,q+1} \to Cl_{p+1,q+1}$ 
because $(J\Delta \gamma_1 g_2)^2=-1$. 
We then find that the associated classifying space is $R_{p-q+1}$.
Thus the conditions for the existence of both $\Delta$ and $g_2$
are given by $\pi_0(R_{p-q})=\pi_0(R_{p-q+1})=0$,
which is met when $q-p=2,3$.
At $q-p=3$, the original classification is trivial.
The case of $q-p=2$ is exactly the second descendant $\mathbb{Z}_2$ 
marked in Table \ref{tab-reflection} as ``$\mathbb{Z}_2$''.

So far we have discussed classes with both TRS and PHS.
Symmetry classes with either TRS or PHS
can be discussed in the same way.
For classes with PHS and $R^-$, the
above discussion is applicable if we drop $e_3$ from the algebras.
For classes with TRS and $R^-$,
while constructions of algebras are slightly different,
the extension problems turn out to have the same structures and 
the resulting existence conditions become the same as the $R^{--}$ case.
Thus, we can conclude that the deformation only occurs
for the second descendant ``$\mathbb{Z}_2$'',
when we have reflection symmetries $R^-$ or $R^{--}$.

\subsection{$R^{++}$ case}
The original classification for the mass term $e_0$ is
obtained from the extension $Cl_{p,q} \to Cl_{p,q+1}$
with the extended algebra $Cl_{p,q+1}$ generated by
\begin{align}
\{e_0,e_1,e_2,e_3,J\gamma_1,\ldots,J\gamma_d, J \gamma_1 R^{++} \},
\end{align}
whose classifying space is $R_{q-p}$.
Let us examine whether the $\Delta$-type operator can be
added to the algebra as
\begin{align}
\{e_0,e_1,e_2,e_3,J\gamma_1,\ldots,J\gamma_d, J \gamma_1 R^{++}, \Delta R^{++} \}.
\end{align}
The existence condition for $\Delta$ is found from the
extension
$Cl_{p,q} \to Cl_{p,q+1}$, for $(\Delta R^{++})^2=+1$.
We note that this is the same extension problem as that for
the classification of $e_0$.
This implies that,
when we have a topologically non-trivial classification for $e_0$,
which is the case of our interest,
we cannot have a $\Delta$-type operator.
Thus deformation of a topologically nontrivial state
to a trivial one is impossible for the $R^{++}$ reflection symmetry.

For classes with either TRS or PHS and with a reflection symmetry $R^+$,
we can repeat similar discussions 
to show that the deformation does not occur.

\subsection{ $R^{-+}$ and $R^{+-}$ case}
The Clifford algebra with the reflection symmetry $R^{-+}$ or $R^{+-}$
is written as
\begin{align}
\{e_0,e_1,e_2,e_3,J\gamma_1,\ldots,J\gamma_d \}
\tensor \{ \widetilde M \},
\end{align}
with $\widetilde M$
that squares to either $+1$ or $-1$.
$\widetilde M$ is given in Table \ref{tab-M-real}(b) 
with $M=J\gamma_1 R$,
according to which $\widetilde M$
is either of $TC\gamma_1 R$ or $TCJ\gamma_1 R$ 
depending on the symmetry class and the type of $R$.

First let us consider the case where $\widetilde M^2=+1$.
We look into the existence condition for $\Delta$ with the algebra
\begin{align}
\{e_0,e_1,e_2,e_3,J\gamma_1,\ldots,J\gamma_d \}
\tensor \{ \widetilde M, J\gamma_1 \Delta \widetilde M \}.
\end{align}
Since $\widetilde M^2=(J\gamma_1 \Delta \widetilde M)^2=+1$,
the existence of $\Delta$ is determined from the extension 
$Cl_{p,q+1} \tensor Cl_{0,0} \to Cl_{p,q+1} \tensor Cl_{0,1}$.
This corresponds to setting $\tilde q=0$ in
Table \ref{tab-multiple}, and we find that
the classification is always $\mathbb{Z}$
(or $\mathbb{Z}\times\mathbb{Z}$).
Thus, non-trivial states cannot be deformed to a trivial one
due to the lack of a $\Delta$-type operator.

When $\widetilde M^2=-1$,
we consider extending an algebra with 
$\Delta$ and $g_2$ to
\begin{align}
\{e_0,e_1,e_2,e_3,J\gamma_1,\ldots,J\gamma_d, J\gamma_1 \Delta g_2 \}
\tensor \{ \widetilde M, J\gamma_1 \Delta\}.
\end{align}
Since $\widetilde M^2=(J\gamma_1 \Delta)^2=-1$,
the existence of $\Delta$ is associated with the extension problem
$Cl_{p,q+1} \tensor Cl_{0,0} \to Cl_{p,q+1} \tensor Cl_{1,0}$,
or equivalently,
$Cl_{p,q+1} \tensor Cl_{0,2} \to Cl_{p,q+1} \tensor Cl_{0,3}$.
By setting $\tilde q=2, m=q+1-p$ in
Eq.~(\ref{extension-multiple-2}) and Table \ref{tab-multiple},
we find that $\Delta$ can exist when $q-p=3,4,5$.
Among these cases, only when $q-p=4$,
the original classification of the Dirac mass $e_0$ is
non-trivial.
We should then look at the existence condition for $g_2$
at $q-p=4$.
Since $(J\gamma_1 \Delta g_2)^2=-1$,
the existence of $g_2$ is associated with
$Cl_{p-1,q+1} \tensor Cl_{2,0} \to Cl_{p,q+1} \tensor Cl_{2,0}$,
or equivalently,
$Cl_{q+3,p-1} \to Cl_{q+3,p}$,
classification of which is $\pi_0(R_0)=\mathbb{Z}$ for $q-p=4$,
and $g_2$ does not exist. 

Thus we conclude that no deformation of non-trivial states
into trivial states can take place in the cases with $R^{+-}$ and $R^{-+}$.

\subsection{Complex cases}
Finally we briefly discuss the absence of the deformation
in the complex classes.
Here we concentrate on the cases when addition of reflection
symmetry changes the topological classification: class AIII with $R^+$
and class A. 
Let us consider the existence condition for a $\Delta$-type
operator which is added to the algebra as
\begin{align}
\{ e_0, ~(e_1,)~ \gamma_1,\ldots, \gamma_d, \gamma_1 R, \Delta R \},
\end{align}
where $e_1=\Gamma$ is present for class AIII but is absent for class A.
Since there is no distinction between generators squaring to
$+1$ and those to $-1$
in the complex Clifford algebra,
we see that the extension problems for classifying $e_0$ and
for the existence of $\Delta$ are the same.
Therefore, if the classification of $e_0$ is non-trivial,
which is the case of our interest, 
then $\Delta$-type operators do not exist,
and no deformation into a trivial state occurs.

\section{Classification for multiple symmetries\label{appendix-mult}}
We briefly explain how to derive classification of time-reversal
invariant topological superconductors
in the presence of multiple additional symmetries.
For systems with both time-reversal and particle-hole symmetries,
the extension problem is given as Eq.\ (\ref{extension-multiple}),
which is equivalent to
\begin{align}
Cl_{0, \tilde q} \tensor Cl_{0,m} \to Cl_{0, \tilde q+1} \tensor Cl_{0,m}.
\end{align}
This equivalence can be understood by using Eq.\ (\ref{p+1,q+1})
several times.

For each value of $m$, 
we make use of the following relations:
\begin{subequations}
\begin{align} 
Cl_{0, \tilde q} \tensor Cl_{0,1}
 &\simeq Cl_{0, \tilde q} \oplus Cl_{0, \tilde q}, \\
Cl_{0, \tilde q} \tensor Cl_{0,2}
 &\simeq Cl_{0, \tilde q}\tensor\mathbb{R}(2), \\
Cl_{0, \tilde q} \tensor Cl_{0,3}
 &\simeq Cl_{0, \tilde q} \tensor Cl_{1,0} \tensor Cl_{0,2}
 \simeq Cl_{\tilde q} \tensor \mathbb{R}(2), \\
Cl_{0, \tilde q} \tensor Cl_{0,4} &\simeq Cl_{0, \tilde q+4}, \\
Cl_{0, \tilde q} \tensor Cl_{0,5}
 &\simeq Cl_{0, \tilde q+4}\tensor Cl_{0,1}
  \simeq  Cl_{0, \tilde q+4} \oplus Cl_{0, \tilde q+4}, \\
Cl_{0, \tilde q} \tensor Cl_{0,6}
 &\simeq Cl_{0, \tilde q+4}\tensor Cl_{0,2}
  \simeq Cl_{0, \tilde q+4}\tensor\mathbb{R}(2), \\
Cl_{0, \tilde q} \tensor Cl_{0,7}
 &\simeq Cl_{0, \tilde q+4}\tensor Cl_{0,3}
  \simeq Cl_{\tilde q+4}\tensor\mathbb{R}(2),
\end{align}
\end{subequations}
which can be derived using Eqs.\ (\ref{eq-app-Cl-R}).
Having in mind that $Cl_{0, q} \to Cl_{0, q+1}$ is classified with $R_q$
and that $Cl_q \to Cl_{q+1}$ is classified with $C_q$,
we obtain the classifying space listed in the last column
in Table \ref{tab-multiple}.


\begin{thebibliography}{30}
\bibitem{note}
For non-minimal (or reducible) models, we have the following counter example.
Let us assume that operators $e_1$, $\ldots$, $e_{p+q-1}$ in a given
non-minimal Dirac model form a Clifford algebra $Cl_{p,q-1}$.
We further assume that the model has three additional operators
$g_1$, $g_2$, $g_3$ which anticommute with all $e_i$'s and
obey the following conditions:
$[g_1,g_2]=[g_1,g_3]=\{g_2,g_3\}=0$ and $g_1^2=g_2^2=g_3^2=+1$.
In this case the product $g_1g_2$ is a constant of motion,
and the Hamiltonian can be block diagonalized and therefore is non-minimal.
We can extend the Clifford algebra $Cl_{p,q-1}$ to $Cl_{p,q}$
by adding $g_1$ as a unique allowed generator, and
topology of the associated classifying space is either $\mathbb{Z}$
or $\mathbb{Z}_2$.
However, it is also possible to 
extend  $Cl_{p,q}$ generated by $\{e_1,\ldots, e_{p+q-1}, g_2 \}$,
to $Cl_{p,q+1}$ generated by $\{e_1,\ldots, e_{p+q-1}, g_2, g_3 \}$.
This situation does not happen for a (fully-diagonalized) minimal model,
since we cannot have operators such as $g_2$ or $g_3$
which leads to a block diagonalization with respect to $g_1g_2$ or $g_1g_3$.
\end{thebibliography}

\end{document}